\documentclass[aip,rsi,graphicx,twocolumn]{revtex4-1}
\pdfoutput=1

\usepackage{upgreek}
\usepackage{chemmacros}
\usepackage{siunitx}
\usepackage{setspace}

\newcommand{\ASTRID}{{\small ASTRID}2}

\newcommand{\AMOLine}{{\small AMOL}ine}

\newcommand{\EUV}{{\small EUV}}
\newcommand{\FWHM}{{\small FWHM}}
\newcommand{\RMS}{{\small RMS}}
\newcommand{\HND}{{\small HND}}
\newcommand{\ICD}{{\small ICD}}
\newcommand{\KE}{{\small KE}}
\newcommand{\VMI}{{\small VMI}}
\newcommand{\SMI}{{\small SMI}}
\newcommand{\MEVELER}{{\small MEVELER}}
\newcommand{\MCP}{{\small MCP}}
\newcommand{\MCTWO}{{\small MC}2}
\newcommand{\TOF}{{\small TOF}}
\newcommand{\HEX}{{\small HEX75}}
\newcommand{\OAR}{{\small OAR}}
\newcommand{\PEPICO}{{\small PEPICO}}
\newcommand{\QMS}{{\small QMS}}
\newcommand{\LT}{{\small LT}}
\newcommand{\SIMION}{{\small SIMION}}
\newcommand{\TDC}{{\small TDC}}
\newcommand{\TTL}{{\small TTL}}
\newcommand{\TCPIP}{{\small TCP/IP}}
\newcommand{\XUV}{{\small XUV}}
\newcommand{\UV}{{\small UV}}
\newcommand{\IR}{{\small IR}}

\begin{document}
\singlespacing


\title{A new endstation for extreme-ultraviolet spectroscopy of free clusters and nanodroplets}



\author{Bj\"{o}rn Bastian}
\author{Jakob D. Asmussen}
\author{Ltaief Ben Ltaief}
\affiliation{Department of Physics and Astronomy,
  Aarhus University, Ny Munkegade 120, 8000 Aarhus C, Denmark}
\author{Achim Czasch}
\affiliation{Institut für Kernphysik, Goethe Universität,
  Max-von-Laue-Strasse 1, 60438 Frankfurt, Germany}
\author{Nykola C. Jones}
\author{S{\o}ren V. Hoffmann}
\author{Henrik B. Pedersen}
\author{Marcel Mudrich}
\email[]{mudrich@phys.au.dk}
\affiliation{Department of Physics and Astronomy,
  Aarhus University, Ny Munkegade 120, 8000 Aarhus C, Denmark}


\date{\today}

\begin{abstract}
  We present a new endstation for the \AMOLine{} of the \ASTRID{} synchrotron at
  Aarhus University, which combines a cluster and nanodroplet beam source with a
  velocity map imaging and time-of-flight spectrometer for coincidence imaging spectroscopy.
  Extreme-ultraviolet spectroscopy of free nanoparticles is a powerful tool
  for studying the photophysics and photochemistry of resonantly excited or ionized
  nanometer-sized condensed-phase systems. Here we demonstrate this capability by
  performing photoelectron-photoion coincidence (\PEPICO{})
  experiments with pure and doped superfluid helium
  nanodroplets.
  Different doping options and beam sources provide a versatile
  platform to generate various van der Waals clusters as well
  as \ch{He} nanodroplets.
  We present a detailed characterization of the new setup and present examples
  of its use for measuring high-resolution yield spectra of charged particles, time-of-flight
  ion mass spectra, anion--cation coincidence spectra, multi-coincidence electron
  spectra and angular distributions.
  A particular focus of the research with this new endstation is on
  intermolecular charge and energy-transfer processes in heterogeneous
  nanosystems induced by valence-shell excitation and ionization.
\end{abstract}


\maketitle 

\section{Introduction}
\enlargethispage{-\baselineskip}

Synchrotron radiation has been used to study the structure and dynamics of
atoms, molecules and clusters since the 1980s. Nevertheless, synchrotron
spectroscopy of gas-phase targets still is an active research field today, due
to the advantages of synchrotrons over lasers and other light sources: The
enormous range of photon energies covered by synchrotrons, spanning from the
infrared up to hard X-rays, is unchallenged by
any other radiation source. State-of-the-art monochromator beamlines typically
provide extremely narrow-band radiation ($E/\Delta E > 10^4$) continuously
tunable across wide spectral ranges. This allows us to perform high-resolution
spectroscopy in the extreme-ultraviolet (\EUV{}) or X-ray ranges, not accessible by
other means.

Over the past decades, the focus of gas-phase synchrotron science has been
shifting from mere atomic and molecular structure determination to unraveling
the complex photophysics and photochemistry of free biomolecules and nanoparticles.
Atomic and molecular clusters are ideal model systems with reduced complexity that
enable detailed studies of relaxation dynamics in condensed phase systems initiated
by valence-shell or inner-shell excitation or ionization. In particular,
interatomic or intermolecular charge and energy transfer are fundamentally
important relaxation processes of weakly bound molecules and nanosystems exposed
to ultraviolet up to X-ray radiation. Nanoparticles are of great
interest for modern applications due to their peculiar properties determined by
finite-size effects; applications range from heterogeneous
photocatalysis~\cite{Han:2018}, nanoplasmonics~\cite{liz2014nanoplasmonics}, aerosol
science~\cite{wang2005nanoparticle}, to radiation biology~\cite{butterworth2012physical}.

Recent technical developments in synchrotron technology have mostly aimed
to increase the achievable photon energy, brilliance and coherence of
X-ray beams. In the particular field of gas-phase studies at synchrotrons, major
progress has been made by devising more and more sophisticated diagnostic
methods. Imaging detection of ions and electrons emitted by the irradiated
molecules and nanoparticles provides both information about magnitudes and angles
of electron and ion momenta. In particular, the velocity map imaging (\VMI{})
technique~\cite{Eppink1997:rsi} is now widely used at \EUV{} gas-phase beamlines
because of its high particle momentum resolution and its $4\pi$-collection
efficiency. Additionally, the combination of \VMI{} of electrons with
time-of-flight (\TOF{}) detection of ions, or vice versa, or even double-sided
\VMI{} detection, where the electrons and ions are detected in coincidence
(photoelectron-photoion coincidence, \PEPICO{}), are established techniques at many
facilities nowadays~\cite{garcia2005refocusing, hosaka2006coincidence,
  rolles2007velocity, tang2009threshold, o2011photoelectron,
  garcia2013delicious, bodi2012new, tang2015vacuum, kostko2017soft}.

Gas-phase synchrotron science is also being moved forward by devising and implementing
more diverse sources of free molecular complexes, clusters and
nanoparticles into \EUV{} and X-ray beamlines. While biologically
relevant molecules are mostly thermally evaporated~\cite{prince2015study},
supersonic jet expansions are also used, where the carrier gas is bubbled
through water and passed over the sample vapor to form microhydrated
clusters~\cite{khistyaev2013proton}. A slow-flow laser photolysis
reactor in the vicinity of the ionization region is used to measure unimolecular and
bimolecular reactions of free radicals~\cite{sztaray2017crf}. Using the
photon-ion merged-beams technique~\cite{schippers2016photoionisation},
highly charged ions~\cite{simon2010photoionization}, size-selected metal
clusters~\cite{niemeyer2012spin}, and endohedral fullerenes have been
investigated~\cite{kilcoyne2010d}. By coupling an aerosol source to a synchrotron beamline,
particles up to a size of $\sim \SI{1}{\micro\meter}$ can be studied by \EUV{}
radiation~\cite{shu2006coupling,gaie2011vuv,goldmann2015electron}.

Another attractive target system for \EUV{} synchrotron studies is pure and doped
superfluid helium (\ch{He}) nanodroplets
(\HND{}s)~\cite{Toennies2004:acie,mudrich2014photoionisaton}.
Pioneering work by the groups of Toennies~\cite{frochtenicht1996photoionization},
M{\"o}ller~\cite{von1997discrete}, and Neumark~\cite{ziemkiewicz2015ultrafast},
have revealed complex relaxation dynamics of resonantly excited \ch{He}
nanodroplets, including visible and \EUV{} fluorescence
emission~\cite{von1997discrete}, autoionization causing the emission of
ultraslow electrons~\cite{peterka2003photoelectron}, the ejection of mostly
\ch{He2+} ions, and, to a lesser extent, larger fragment ions.
Dopant molecules embedded inside the \ch{He} droplets were found to be efficiently
ionized indirectly by Penning-like excitation-transfer ionization via \ch{He^*}
`excitons'.\cite{frochtenicht1996photoionization, wang2008photoelectron,
  peterka2006photoionization}

Neumark \textit{et al.}\ introduced the \VMI{} technique to study \ch{He}
nanodroplets in the photon energy range around the ionization energy of \ch{He},
$E_i^{\ch{He}}=\SI{24.59}{eV}$, using synchrotron radiation and later using laser-based \EUV{}
sources~\cite{ziemkiewicz2015ultrafast}. In recent experiments at the
synchrotron facility Elettra, Trieste, we have extended that work by
implementing photoelectron–photoion coincidence (\PEPICO{}) \VMI{} detection. In this
way, we can measure photoelectron spectra and angular distributions in
coincidence with specific ion masses~\cite{buchta2013charge}. We have evidenced efficient single
and double ionization of alkali-metal atoms and molecules attached to the
surface of \ch{He} droplets, through resonantly excited \ch{He}
nanodroplets~\cite{buchta2013charge,ben2019charge,laforge2019highly}.
Upon photoionization of the droplets, we found that embedded metal atoms are
ionized by charge transfer mechanisms~\cite{buchta2013charge, laforge2019highly,
  ltaief2020electron}.
We have recently complemented these synchrotron studies by time-resolved
experiments using the tunable \EUV{} free-electron laser Fermi,
Trieste~\cite{mudrich2020ultrafast, laforge2021ultrafast, Asmussen2021,asmussen2022time}.

\begin{figure*}[ht]
  \includegraphics[width=\linewidth]{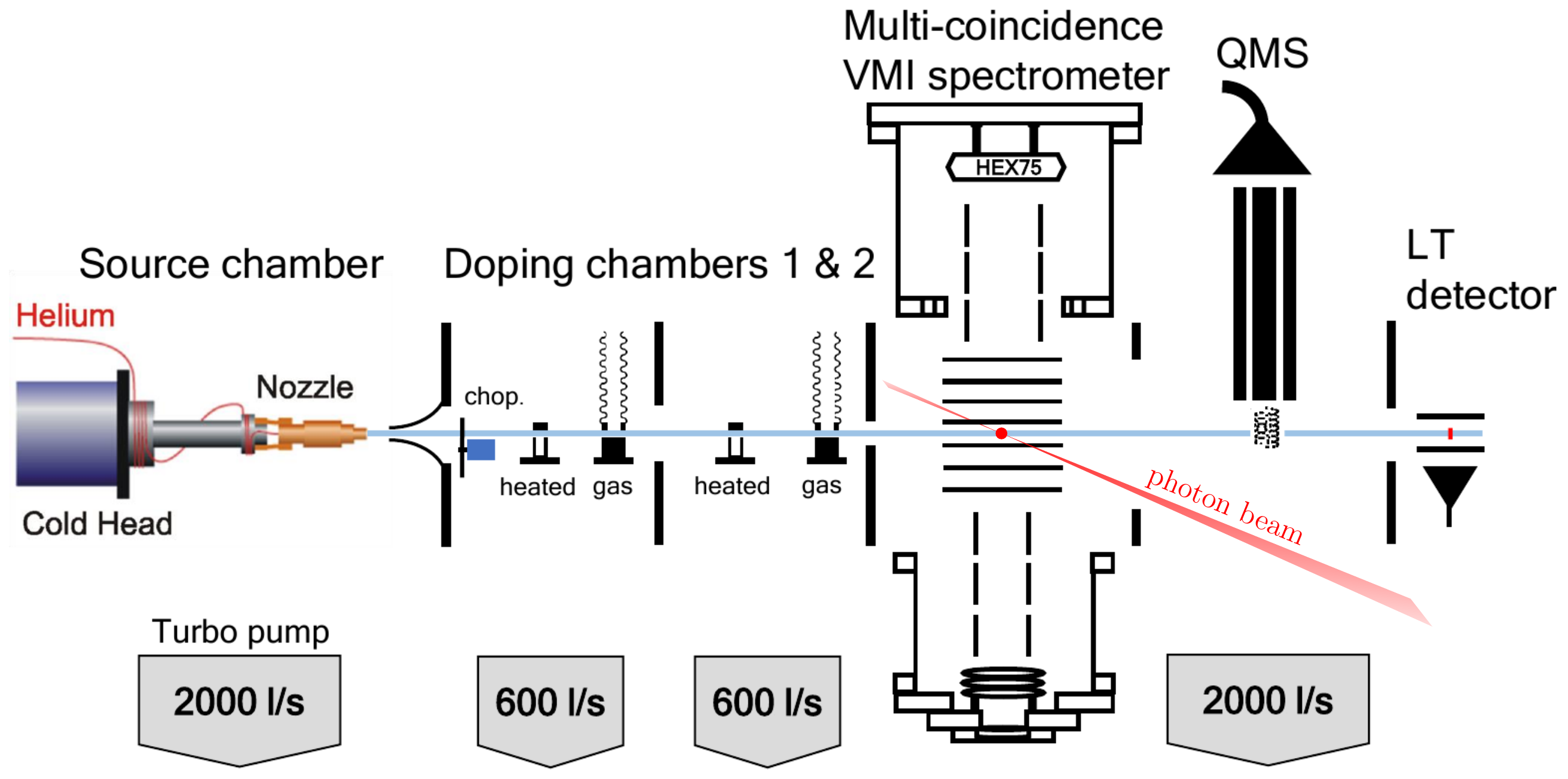}
  \caption{\label{fig:setup}%
    Experimental setup. A skimmed \ch{He} droplet beam is produced from a
    cryogenically cooled continuous nozzle expansion with \SI{5}{\micro\meter} diameter.
    The chopper wheel (`chop.') allows to alternatively measure foreground and
    background.
    Gas doping and oven cells enable pickup of dopants by the droplets.
    The droplets then cross the photon beam in the center of the spectrometer.
    A quadrupole mass spectrometer (\QMS{}) and Langmuir-Taylor detector (\LT{})
    behind the spectrometer are used for beam diagnostics.
  }
\end{figure*}

In this article we describe a new endstation that has recently been installed at
the \AMOLine{} of the \ASTRID{} synchrotron at Aarhus University,
Denmark~\cite{hertel2011astrid2}. With this apparatus, our previous experiments
at Elettra~\cite{buchta2013charge, laforge2019highly} will be further extended
owing to a more sophisticated imaging detector, extended doping capabilities,
improved options for cluster beam characterization, and various cluster sources.
We characterize the new \PEPICO{} \VMI{} spectrometer and present original data that
illustrate the capability of the new setup to reveal the details of interatomic
Coulombic decay (\ICD{}) processes occurring in pure and doped \ch{He}
nanodroplets.\cite{Shcherbinin2017:pra}

\section{Experimental setup}

The new endstation placed at the end of the \AMOLine{} of the \ASTRID{}
synchrotron is schematically shown in Fig.~\ref{fig:setup}.
A skimmed atomic, molecular or cluster beam traverses two vacuum chambers
for doping with gas and oven cells before crossing the photon beam from the
synchrotron in the center of an ion--electron coincidence spectrometer.
The molecular or cluster beam then passes a chamber housing a quadrupole mass
spectrometer (\QMS{}) and is dumped in an additional small chamber, which is
directly attached to the adjacent chamber without additional pumping. This beam
dump also houses a simple Langmuir-Taylor (\LT{}) detector for alkali and
earth-alkali metals.
Both chambers serve the purpose of monitoring the intensity and composition of
the cluster beams. The pressure rise in the \LT{} chamber in the
presence of an \ch{He} atomic or nanodroplet beam is on the order of
\SI{e-8}{mbar}.

For generating \ch{He} nanodroplets we use a continuous cryogenic nozzle of
diameter \SI{5}{\micro\meter} placed at a distance of \SI{12}{mm} in front of a
skimmer with \SI{0.4}{mm} diameter.
Unless otherwise stated, all presented data were recorded with \SI{30}{bar}
\ch{He} backing pressure and \SI{14}{K} nozzle temperature.
The average number of \ch{He} atoms per droplet was estimated by titration
measurements~\cite{Gomez2011:jcp} that yielded
$\langle N_{\ch{He}} \rangle \approx 2000$.
Typical operating pressures under these conditions
are \SI{1.4e-3}{mbar} in the source, \SI{4e-6}{mbar} in the first
and \SI{6e-7}{mbar} in the second doping chamber, \SI{3e-8}{mbar}
in the \VMI{} and \QMS{} chambers and \SI{5e-8}{mbar} in the \LT{} chamber.

\begin{figure*}[ht]
  \includegraphics[width=\linewidth,trim=60 435 2 0,clip]{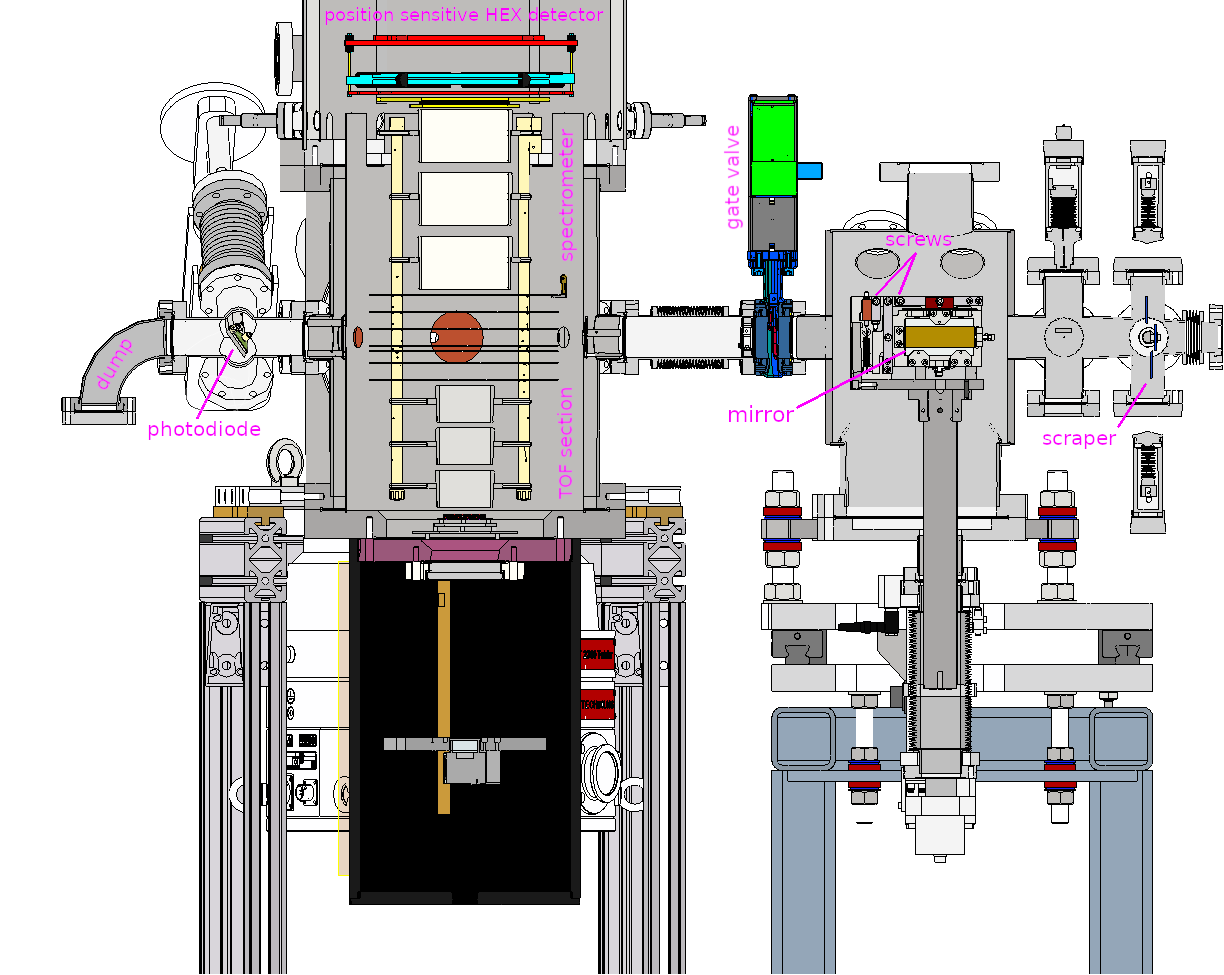}
  \caption{\label{fig:setup:mirror}%
    Spectrometer with beam dump and photodiode (left) and grazing
    incidence mirror and scrapers (right) for the incoming photon beam.
  }
\end{figure*}

\subsection{The \AMOLine{} at \ASTRID{}}
\enlargethispage{\baselineskip}

A cross-section of the spectrometer chamber and the adjacent mirror chamber to
couple in the photon beam from the \AMOLine{} is presented in
Fig.~\ref{fig:setup:mirror}. Two different gratings for the monochromator
(not shown)
provide 5 to \SI{150}{eV} photons from an undulator source
on the \ASTRID{}
synchrotron. A slit and a selection of different foils
can be used to adjust the photon intensity and suppress
higher-order radiation. A focusing mirror after the monochromator is used to align
the photon beam on the mirror (\MCTWO{}) in Fig.~\ref{fig:setup:mirror} that refocuses the
beam into the interaction region of the spectrometer
(see Supporting Information S1).
Several sets of scrapers placed upstream in the beamline
are instrumental for reducing stray light reaching the mirror.
The vertical position of
the mirror is electronically
controlled and can be replaced with either a photodiode or phosphor screen for
photon beam diagnostics. The horizontal position (perpendicular to the
beam direction) can be adjusted by movement of the chamber, while the pitch and yaw of the mirror are
manually adjusted by a set of in-vacuum screws.
The mirror chamber is separated from the spectrometer chamber
by a pneumatic gate valve and a pinhole. The photon beam is dumped in a bent
tube on the opposite side of the spectrometer which contains another phosphor
screen and photodiode for beam diagnostics.
Software integration of the data acquisition program and beamline control via
\TCPIP{} allows several beamline parameters to be scanned for optimization,
in particular the photon energy for automated photon-energy scans.

\subsection{Beam source and doping options}

A \ch{He} atomic or nanodroplet beam is formed by a precooled continuous supersonic
expansion through a \si{\micro\meter}-size nozzle. The nozzle holder, body and
\ch{He} gas line are cooled by a closed-cycle cryostat that has a cooling power of
\SI{1}{W} at \SI{4}{K}.
A copper heat shield mounted on the first cooling stage is not shown in
Fig.~\ref{fig:setup}.
The nozzle temperature is continuously measured and regulated by resistive
heating. Nozzle openings are formed by planar Pt foils (orifice diameters 5, 10
or \SI{20}{\micro\meter}) pressed into a copper holder. The open cylinder at the
back of this holder is pressed onto a copper counterpart by a union nut and
sealed with a C-shaped sealing ring. This facilitates
the exchange of nozzles with different diameter. New nozzles are tested by
measuring the \ch{He} flow at \SI{50}{bar} backing pressure, which typically amounts to
\SI{1}{\milli\litre\per\second} for a \SI{5}{\micro\meter} nozzle.
The supersonic expansion is skimmed to select the cold part of the beam before a
shock wave is formed and for ensuring differential pumping.
The skimmer holder is screwed into a threaded outer surface to adjust the
nozzle-to-skimmer distance.
The horizontal and vertical nozzle position can be varied by a translation
stage without breaking vacuum.

To expand other gases, the same \ch{He} nanodroplet source can be used where only the nozzle part has to
be replaced by one with a larger diameter. Alternatively, it can be fully replaced by another beam
source. A source for pure or molecule-doped water clusters based on the design of
Förstel~\textit{et al.}~\cite{Forstel:2015} is currently in preparation.

In the oven chambers, four slots for heated ovens or gas doping cells enable sequential pickup of
different dopants into \ch{He} nanodroplets from solid, liquid or gaseous samples.
This allows us to quickly apply different dopants or to form tailored
multicomponent structures by adjusting the partial pressures and the order of
doping the various substances into the droplet beam.
The doping conditions are typically characterized by the \QMS{} or
by charge transfer ionization of dopants when irradiating the droplets with photons at
energies $h\nu\geq E_i^{\ch{He}}$.

\begin{figure*}[ht]
\includegraphics[width=\linewidth]{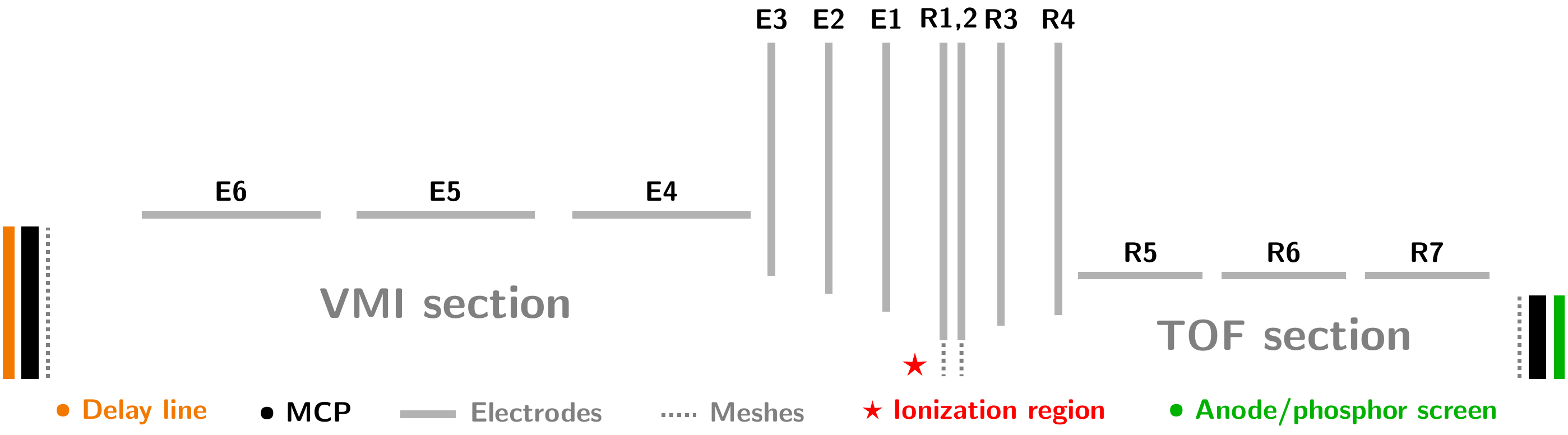}%
\caption{\label{fig:spectrometer}Spectrometer geometry.
  Half of the central cross-section of the electrode stack is depicted true
  to scale, including the center of the ionization region on the central axis.
  The inner electrodes are realized by metal plates with concentric holes. The
  outer ion stacks in Einzel lens configuration are realized by metal cylinders.
  The black bars on the far left and right specify the detector positions.
  The lengths of the \VMI{} and \TOF{} sections from the ionization region to the
  detectors are \SI{247}{mm} and \SI{174}{mm}, respectively.
  Electrode positions and diameters are specified in
  Table~\ref{tab:spectrometer}.
}%
\end{figure*}

\begin{table*}
  \caption{\label{tab:spectrometer}%
    Electrode positions and inner radii of the spectrometer in
    Fig.~\ref{fig:spectrometer}.
  }
  \footnotesize
  \begin{tabular}{l
      @{~\,}r@{.}l
      @{~\,}r@{.}l
      @{~\,}r@{.}l
      @{~\,}r@{.}l
      @{~\,}r@{.}l
      @{~\,}r@{.}l
      @{~\,}r@{.}l
      @{~\,}r@{.}l
      @{~\,}r@{.}l
      @{~\,}r@{.}l
      @{~\,}r@{.}l
      @{~\,}r@{.}l
      @{~\,}r@{.}l
    }
    \toprule
    Electrode
    & \multicolumn{2}{c}{E6}
    & \multicolumn{2}{c}{E5}
    & \multicolumn{2}{c}{E4}
    & \multicolumn{2}{c}{E3}
    & \multicolumn{2}{c}{E2}
    & \multicolumn{2}{c}{E1}
    & \multicolumn{2}{c}{R1}
    & \multicolumn{2}{c}{R2}
    & \multicolumn{2}{c}{R3}
    & \multicolumn{2}{c}{R4}
    & \multicolumn{2}{c}{R5}
    & \multicolumn{2}{c}{R6}
    & \multicolumn{2}{c}{R7}
    \\
    \hline
    Position/\si{mm}
    & -190 & 625 
    & -130 & 875 
    & -70 & 625  
    & -40 & 0    
    & -24 & 0    
    & -8 & 0     
    & 8 & 0      
    & 13 & 0     
    & 24 & 0     
    & 40 & 0     
    & 62 & 875   
    & 102 & 875  
    & 142 & 875  
    \\
    Radius/\si{mm}
    & 42 & 0
    & 42 & 0
    & 42 & 0
    & 25 & 0
    & 20 & 0
    & 15 & 0
    & 7 & 0
    & 7 & 0
    & 11 & 0
    & 14 & 0
    & 25 & 0
    & 25 & 0
    & 25 & 0
    \\
    \hline
  \end{tabular}
\end{table*}

\subsection{Spectrometer and \VMI{} potentials}

Two microchannel plate (\MCP{}) detectors with a stack of $\mu$-metal shielded
electrodes and a position-sensitive detection system, see
Fig.~\ref{fig:setup:mirror}, are currently set up to measure
kinetic energies on one side and ion flight times for mass spectrometry on the
other side of the spectrometer. Coincidence detection of cations and electrons
allows us to obtain the ion time-of-flight (\TOF{}) as the
difference of electron and ion arrival times on the detectors. This detection mode
is ideal for experiments with quasi-continuous photon beams as provided by synchrotrons.

The key strength of the electron and ion spectrometer is the combination
of a RoentDek hexanode delay-line detector (\HEX{})~\cite{jagutzki2002multiple} with a set of electrodes to
perform high-resolution \VMI{}.
The position-sensitive \HEX{} detector is well suited for single particle
detection at high event rates \cite{Oesterdahl2005:jpcs} and in particular to
reconstruct multi-hit events.
In case of ion imaging, the $<\SI{1}{ns}$ time resolution facilitates the full
three-dimensional reconstruction of the momentum vectors of several ions in
Coulomb explosion experiments~\cite{Hasegawa2001:cpl}.

Fig.~\ref{fig:spectrometer} shows a scaled drawing of the electrode configuration.
The \VMI{} section is similar to the spectrometer geometry at the
\hbox{GasPhase} beamline at Elettra in Trieste, Italy~\cite{o2011photoelectron}.
A third extractor electrode is inserted and the flight tube is split into three
tube sections to be used as an Einzel lens.
The \TOF{} and \VMI{} sections of the spectrometer are separated by two meshes
in the center that serve as repellers for \VMI{}, labeled R1 and
R2 in Fig.~\ref{fig:spectrometer}.
The \TOF{} section has a similar geometry with one electrode less, smaller electrode
radius and shorter length.
The time resolution does not noticeably depend on the specific electrode
potentials, which
are chosen to defocus the ions on the \SI{40}{mm} diameter \MCP{}
to reduce wear (see Figure~\ref{fig:defocus}).
The combination of the \MCP{} with a phosphor screen and digital camera
allows us to monitor the spatial distribution of ions on the detector. When removing the central meshes,
\VMI{} could be performed on both sides of the spectrometer.
The delay-line detector is combined with a larger \SI{80}{mm} diameter \MCP{}
which has the advantage of improved resolution when using higher \VMI{} magnification
and a larger maximum kinetic energy acceptance at a given magnification factor.

To achieve optimal \VMI{} resolution it is crucial to optimize the potentials of the
extractor electrodes to minimize the position spread due to the initial spatial
distribution in the interaction region. The latter is defined by the overlap of
the focused photon beam and the droplet beam, of which the extent in
the horizontal plane is restricted to \SI{4}{mm} by a pair of adjustable scrapers.
The three extractor electrodes and three electrodes in Einzel lens configuration are
used to adjust the imaging properties to different ranges of electron kinetic energy.
Trajectory simulations with the program \SIMION{}~\cite{Simion2008} were used
with a derivative based optimization routine~\cite{Stei2013:jcp} to refine
different sets of electrode potentials for optimal \VMI{} conditions in a
simple and quick way (see also Supporting Information S2). In the original
method, derivatives are evaluated at zero initial kinetic energy. Here, the
method was extended to nonzero values.


To calibrate the spectrometer in spatial map imaging mode, potassium-doped
\ch{He} nanodroplets
were ionized with a Coherent Mira Ti:sapphire laser at continuous \SI{767}{nm}
wavelength and about \SI{0.56}{W} power.
A lens with \SI{175}{mm} focal length was mounted on a translation stage to
focus the light into the center of the interaction region.
By shifting the focus position in the interaction region horizontally and
vertically, we calibrated the positions on the detector and the ion time-of-flight,
respectively~\cite{Wituschek2016:rsi}. Furthermore, a grating-stabilized
diode laser is at our disposal to implement a simple scheme for photoionization
of potassium atoms for precise calibration measurements~\cite{Wituschek2016:rsi}.

\section{Results and Discussion}

\subsection{Velocity mapping and resolution}

Precise velocity map imaging is the basis for accurate measurements of electron
or ion kinetic energy spectra and angular distributions.
This section describes and characterizes different \VMI{} potentials including
scaled potentials to improve the resolution at low kinetic energies in the
\si{eV} range and use of Einzel lens electrodes to achieve
detection up to kinetic energies of \SI{100}{eV}.

\subsubsection{Electron \VMI{} without the Einzel lens\label{sec:default:vmi}}

\begin{figure}
  \includegraphics[scale=0.9]{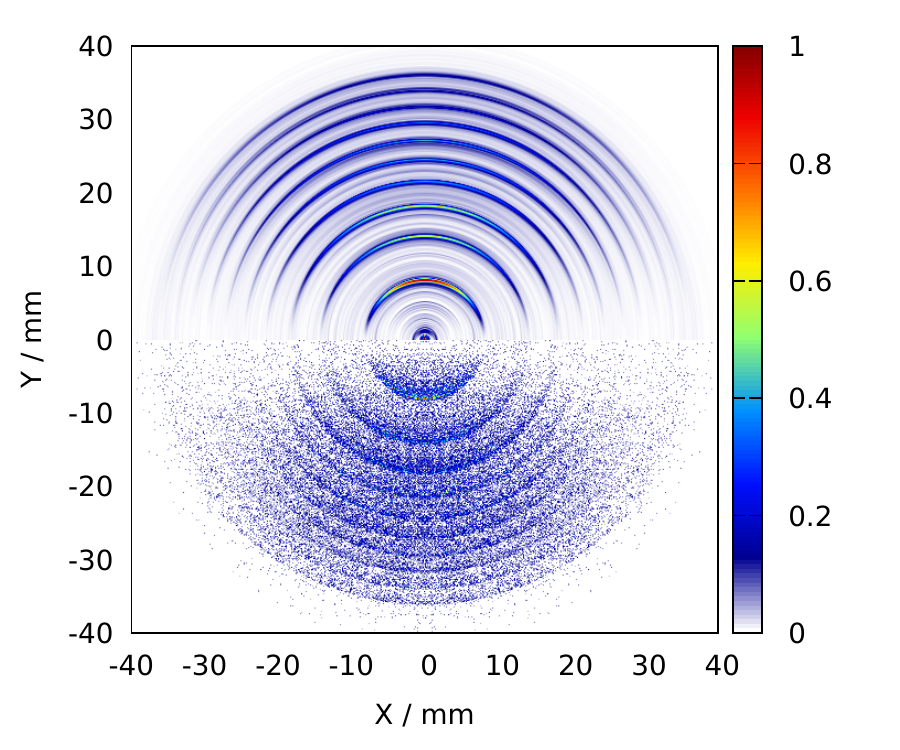}%
  \caption{\label{fig:images}
  Photoelectron images for the ionization of gas-phase \ch{He} at photon energies
  of 26, 29, 32, 35, 38, 41, 44, 47, 50 and \SI{53}{eV}.
  \textbf{Lower half:} Raw images.
  \textbf{Upper half:} Abel inverted distributions.
  }%
\end{figure}

\begin{figure}
  \includegraphics[page=1,width=\linewidth]{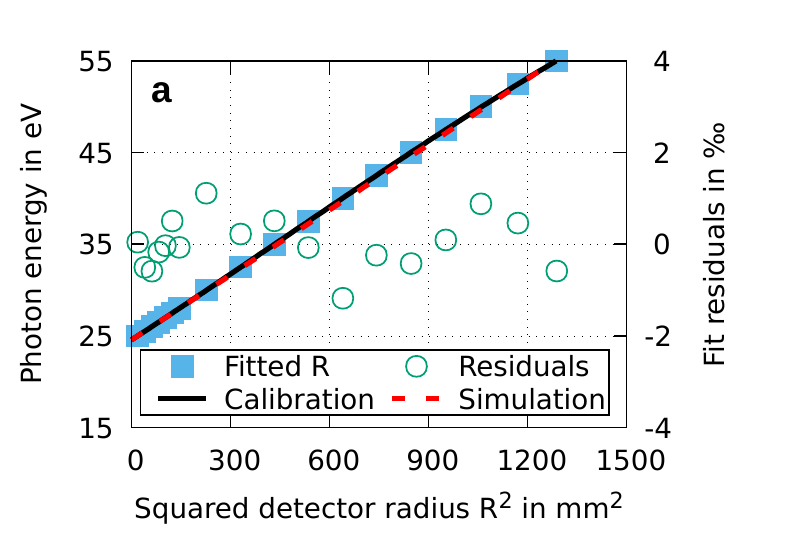}%
  \\%
  \includegraphics[width=\linewidth]{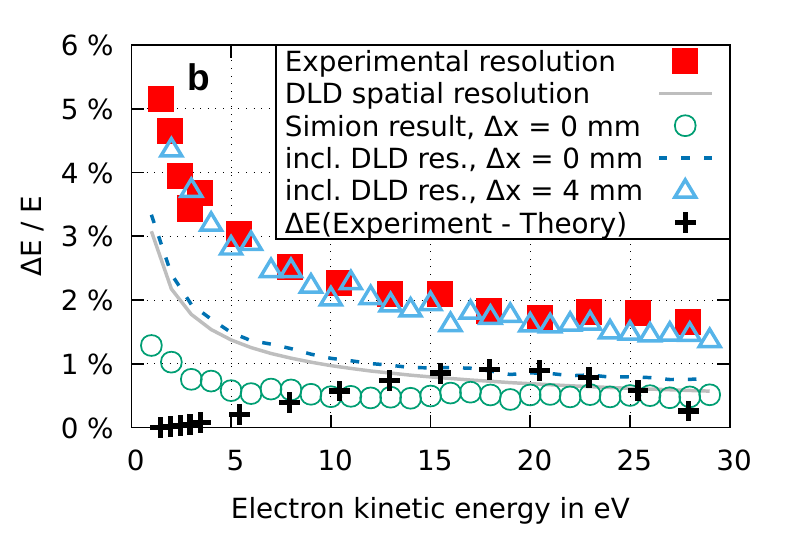}%
  \caption{\label{fig:vmi:calib1}Results of a calibration measurement in comparison with simulation
    results. \textbf{(a)} Photon energy as a function of the squared detector radius
    with a polynomial fit. \textbf{(b)} Experimental energy resolution, simulated
    deviations due to off-center start positions and relative deviation between
    the simulated and the empiric calibration function.
  }%
\end{figure}

Spectrometer potentials for high resolution \VMI{} imaging were determined
with the numerical optimization routine.
We refer to the Supporting Information
for additional details (S2) and the obtained potentials (Table~S.3). 
For calibration and characterization, a set of \ch{He+} images from
photoionization in the gas phase is acquired at equally spaced photon
energies, see Fig.~\ref{fig:images}.
The energy calibration with empirical and simulation results in
Fig.~\ref{fig:vmi:calib1}a and Table~\ref{tab:vmi:calib1} is detailed
in the Supporting Information (S3). 

The agreement of the near-linear
calibration functions is very good with a maximum deviation of
\SI{0.9}{\percent} between simulation and measurement
(black crosses in Fig.~\ref{fig:vmi:calib1}b).
The higher-order coefficients $p_1$ and $p_2$ are not reproduced by the
simulation. Their order of magnitude can be reproduced by a more
detailed simulation taking into account an initial position distribution
(see Supporting Information S4) that is offset from the center by \SI{4}{mm}
away from the detector and \SI{9}{mm} along the photon beam
(last row in Table~\ref{tab:vmi:calib1}).
Besides imperfect centering of the photon beam and the \ch{He} beam, also
small deviations from the ideal electrical field and residual magnetic fields
may contribute to the nonlinearities.

\begin{table}
  \caption{\label{tab:vmi:calib1}%
    Calibration coefficients for \VMI{} without Einzel lens.
    The calibration function is
    $E_\mathrm{kin} = p_0 R^2 + p_1 R^4 + p_2 R^6$.
    Uncertainties in parentheses are given in units of the last digit.
  }
  \begin{tabular}{lccc}
    \toprule
    Origin & $p_0$ & $p_1/\text{\small\num{e-7}}$ & $p_2/\text{\small\num{e-9}}$ \\
    \hline
    Experiment & \num{0.0235+-0.0001} & \num{21+-3} & \num{-1.5+-0.1} \\
    Simulation & \num{0.02358}~\,\, & \num{1.13+-0.02} & \num{0.0013+-0.0008} \\
    ~with offset & \num{0.0260+-0.0004} & \num{26+-11} & \num{-1.5+-1.0} \\
    \hline
  \end{tabular}
\end{table}

\subsubsection{Energy resolution limits}

Quantifying the energy resolution by the ratio of the standard deviation $\Delta E$ to the
magnitude of the electron kinetic energy $E$, the \VMI{} settings without Einzel lens
perform best with $\Delta E/E=\SI{1.7}{\percent}$ at large detector radii.
The resolution stays below \SI{2.1}{\percent} down to about \SI{13}{eV} kinetic
energy in Fig.~\ref{fig:vmi:calib1}b.
Below \SI{13}{eV}, the absolute radial spread stays constant near
\SI{230}{\micro\meter}
so the energy resolution decreases and reaches \SI{5}{\percent} at \SI{1.5}{eV}
kinetic energy.

A position resolution of \SI{160}{\micro\meter}~\cite{Oesterdahl2005:jpcs} \FWHM{}
and better can be achieved with the \HEX{} detector
for atoms and ions,
corresponding to a standard deviation of \SI{68}{\micro\meter}.
Electrons produce signal amplitudes with lower magnitude and higher
variance and we expect a resolution limit of \SI{100}{\micro\meter}
represented by the gray line in Fig.~\ref{fig:vmi:calib1}b.
The simulated energy broadening (see details in Supporting Information S4) due
to the initial spatial distribution in the interaction region (green circles)
only slightly raises this resolution limit (blue dashed line) that stays more
than a factor \num{2} below the experimental resolution.
One possible explanation is the so far not controlled vertical position of
the photon beam:
The assumption of a \SI{4}{mm} offset from center irrespectively of the
direction mostly reproduces the experimental result (blue triangles).
The horizontal position has a minor effect on the simulated resolution.
A later improvement of the detector resolution with higher \MCP{} gain
did not result into a better experimental resolution which supports the
hypothesis that ajdustment of the interplay between vertical position and
electrode potentials may further improve the achieved energy resolution.

Alternative potentials with improved resolution for low and high electron
kinetic energy ranges are presented in the following.

\subsubsection{Alternative \VMI{} potentials\label{sec:alternative:vmi}}

The maximum kinetic energy acceptance with \SI{-3.5}{kV} extraction potential
without using the Einzel lens is about \SI{32}{eV}. In highly excited or core
ionized systems, it is often desirable to capture faster electrons to probe the
process of interest or to include photoelectrons from direct ionization as a
reference, e.g.\ to determine branching ratios.

\begin{figure*}
  \includegraphics{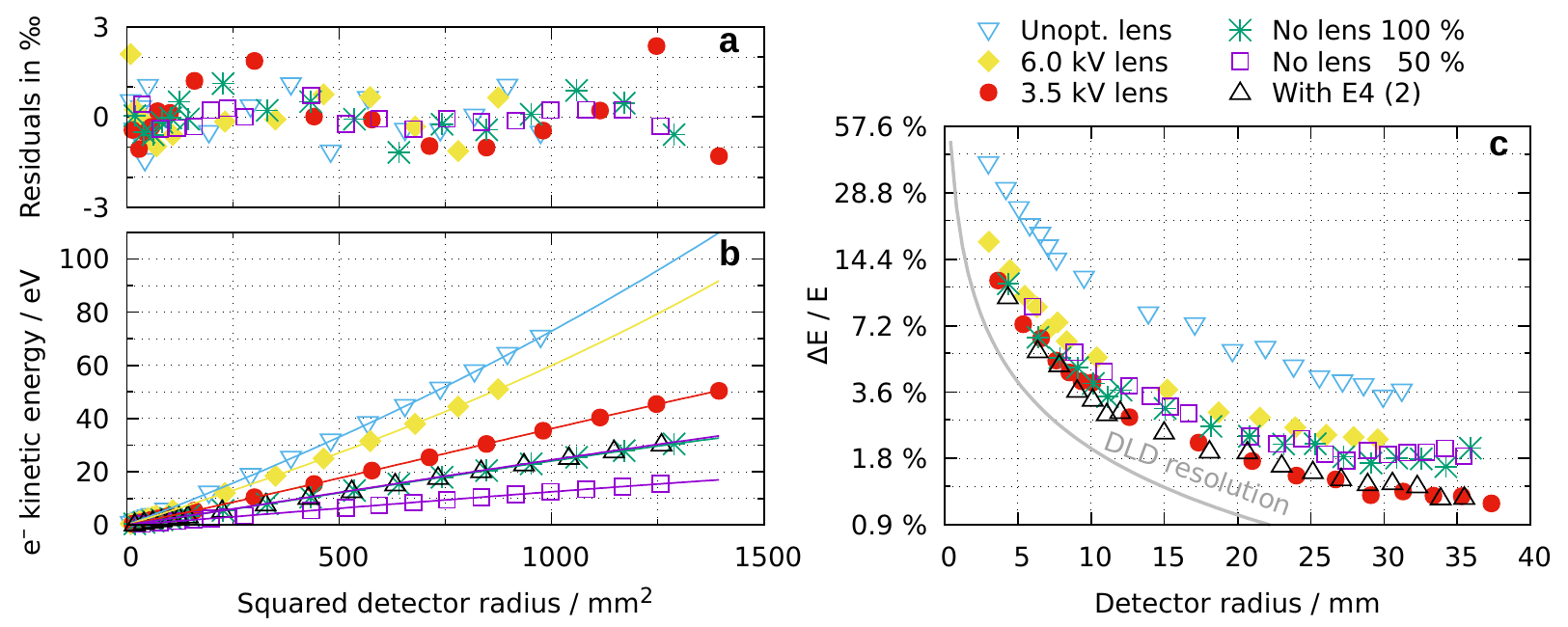}\\%
  \caption{\label{fig:nov:calibration}
    Fit residuals (\textbf{a}), calibration fits (\textbf{b}) and energy
    resolution (\textbf{c}) for different \VMI{} settings. The corresponding
    electrode potentials are specified in Table~S.3. 
    The gray line indicates the expected resolution limit
    of the \HEX{} detector for electrons.
  }%
\end{figure*}

We have achieved compression of electron \VMI{} images with identical positive
potentials on the outer electrodes E4, E6 of the Einzel lens and refer to
the Supporting Information (S2) for additional explanations.
The calibration curves in Fig.~\ref{fig:nov:calibration}b visualize the resulting
energy acceptance. There is no empirical data above \SI{70}{eV} kinetic energy
as the high energy grating for the monochromator was not available when the data
were taken.

Simply adding \SI{6}{kV} potentials to the optimized potentials without the lens
gives the highest energy acceptance up to \SI{100}{eV} but significantly
decreases the energy resolution, see Fig.~\ref{fig:nov:calibration}c
("Unopt.~lens", blue triangles).
Numerical re-optimization for \VMI{} conditions almost restores the original
energy resolution (yellow diamonds) at the cost of somewhat lower \SI{90}{eV}
maximum energy.

Optimized potentials with a \SI{3.5}{kV} Einzel lens are a good compromise with
\SI{50}{eV} energy acceptance and excellent resolution reaching down to
\SI{1.2}{\percent} in the high-energy range.

\begin{figure*}
  \includegraphics{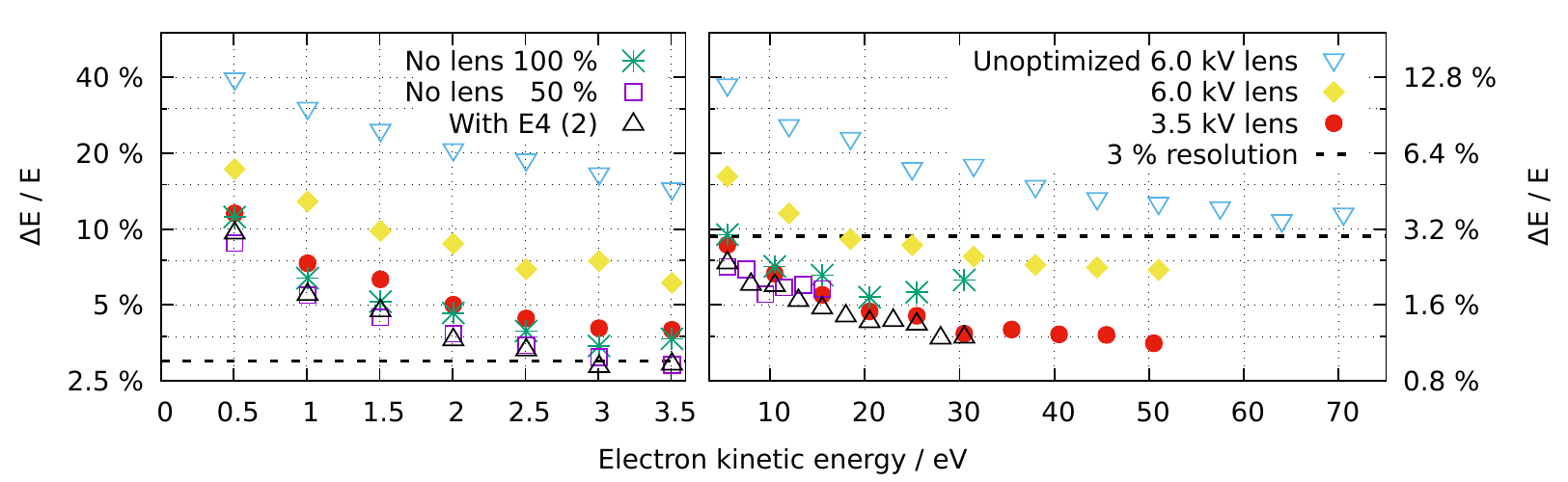}\\%
  \caption{\label{fig:nov:resolution}
    Energy resolution as a function of electron kinetic energy for different
    \VMI{} settings. The corresponding electrode potentials are specified in
    Table~S.3. 
    The \SI{6}{kV} Einzel lens settings are applicable up to \SI{90}{eV} and
    above (see Fig.~\ref{fig:nov:calibration}b) which has not been recorded.
    The $x$ range is split in two parts with different $y$ ranges.
    Dashed lines at \SI{3}{\percent} resolution are shown to guide the eye.
  }%
\end{figure*}

The energy resolution decreases quickly at low energies in the \SI{1}{eV} range
and this would still be true if the spatial resolution limit of the
detector (gray line) could be reached.
Therefore, a larger velocity magnification factor is required to
significantly improve the resolution for slow electrons.
This can to some extent be achieved by scaling the extraction potentials
and is demonstrated for the settings without an Einzel lens
scaled to \SI{100}{\percent}
(stars) and \SI{50}{\percent} (squares) of the full potentials.
While the resolution as a function of radius cannot be substantially altered
(Fig.~\ref{fig:nov:calibration}c), the resolution at a specific energy is
improved. This becomes evident in Fig.~\ref{fig:nov:resolution}, which summarizes
all settings with the resolution as a function of kinetic energy.
The \SI{50}{\percent} results demonstrate the limits of this approach: scaling
the potentials by a factor of \num{2} improved the resolution by merely
\SI{15}{\percent} and settings below \SI{30}{\percent} potentials
were empirically found to be inadequate for imaging.

The \SI{100}{\percent} potentials can be further improved by using lens E4 while
the other two electrodes of the Einzel lens are kept at ground potential.
Settings optimized for \SI{30}{eV} kinetic energy (see Supporting Information
Table~S.3) 
reach the resolution of the \SI{50}{\percent} potentials without compromising
the kinetic energy acceptance in Fig.~\ref{fig:nov:resolution}.
The disadvantage is a significantly reduced time-of-flight resolution due to
a shallower extraction potential ("low res." in Fig.~\ref{fig:massspec}).

Another elegant approach for low kinetic energies is to use a short Einzel lens
in the flight tube to magnify the spatial distribution by inversion~\cite{Offerhaus2001:rsi}.
Ramping up the lens first focuses electrons of different kinetic energy until
reaching inversion and then starts magnifying the inverted electron image.
To be efficient, this approach requires a shorter Einzel lens with smaller
diameter than in our spectrometer geometry. The original geometry would thus
limit the spectrometer to relatively low kinetic energies.
Furthermore, such a geometry reduces the \VMI{} quality with a significant
dependence on the initial particle position in a way that cannot be compensated for
by the numerical optimization routine. A more detailed analysis will be required
to evaluate the trade-off between the different sources of uncertainty.

\subsection{Spatial mapping and beam overlap}

\subsubsection{\SMI{} potentials and calibration\label{sec:smi:pot}}
In analogy to \VMI{} but with inverse logic, the spectrometer can
be used for spatial map imaging (\SMI{}) in the interaction region with high
precision,
irrespectively of the initial velocities of the imaged particles.\cite{Stei2013:jcp}
We have employed two-dimensional \SMI{} for calibration and to characterize the
overlap of the photon and \ch{He} beam in the scattering plane.
The numerical optimization routine was employed to determine
potentials for three different settings that are presented in the Supporting
Information (S5): (1) two-dimensional \SMI{} of radial positions, (2) magnified
two-dimensional \SMI{} and (3) one-dimensional \TOF{}-\SMI{} of axial positions.

Calibration measurements for \SMI{} settings were performed by ionizing
potassium doped \HND{}s at \SI{767}{nm} wavelength.
The magnification factor for ion or electron \SMI{} was found to be
\num{-3.2}, in excellent agreement with the simulation value \num{-3.154}
(\SI{1.5}{\percent} deviation).
With the magnifying settings, a factor of \num{-13} was found
(\SI{3}{\percent} deviation) and the \TOF{}-\SMI{} has a scaling factor of
\SI{64}{\nano\second\per\milli\meter} (\SI{10}{\percent} deviation).

\subsubsection{\SMI{} resolution}

Cation \SMI{} with \ch{K+} ions gives an upper bound standard deviation of
\SI{50}{\micro\meter} for the waist of the focused beam (see Supporting Information,
Table~S.6). 
This corresponds to a radius of \SI{160}{\micro\meter} on the
detector, larger than the expected spatial resolution of the \HEX{}
detector of about \SI{70}{\micro\meter} or lower.
The same width is observed for the magnifying settings, which however
corresponds to \SI{670}{\micro\meter} on the detector.
We assume that this is caused by stronger velocity broadening
in line with the simulations that
predict a significant derivative of the impact position with respect to
the initial ion velocity, see Supporting Information (S5).
The \TOF{}-\SMI{} settings give a similar value of \SI{50}{\micro\meter} for the
vertical direction which corresponds to a standard deviation of \SI{3}{ns} in
the ion flight times
which is larger than the optimum time resolution of the \HEX{} detector
of about \SI{0.5}{ns}.

With electron \SMI{} at full and \SI{50}{\percent} potentials (scaling $s = 1$
and $0.5$), the standard deviations are $w = \SI{80}{\micro\meter}$ and \SI{110}{\micro\meter}.
The simplified assumption $w = w_0 + c/\sqrt{s}$ with beam waist $w_0$ and
velocity dependent broadening thus gives the estimate
$w_0 \approx \SI{7.6}{\micro\meter}$.
The order of magnitude is confirmed by the \SI{8.5}{\micro\meter} focused spot
radius of a Gaussian beam with \SI{5}{mm} radius at the focusing lens.
The experimental conditions are therefore suited to assess the
resolution of about \SI{50}{\micro\meter} (\RMS{}) for ions
that is achieved with the \SMI{} mode.


\subsubsection{Width of the \ch{He} droplet beam}

\begin{figure}
  \includegraphics[width=\linewidth]{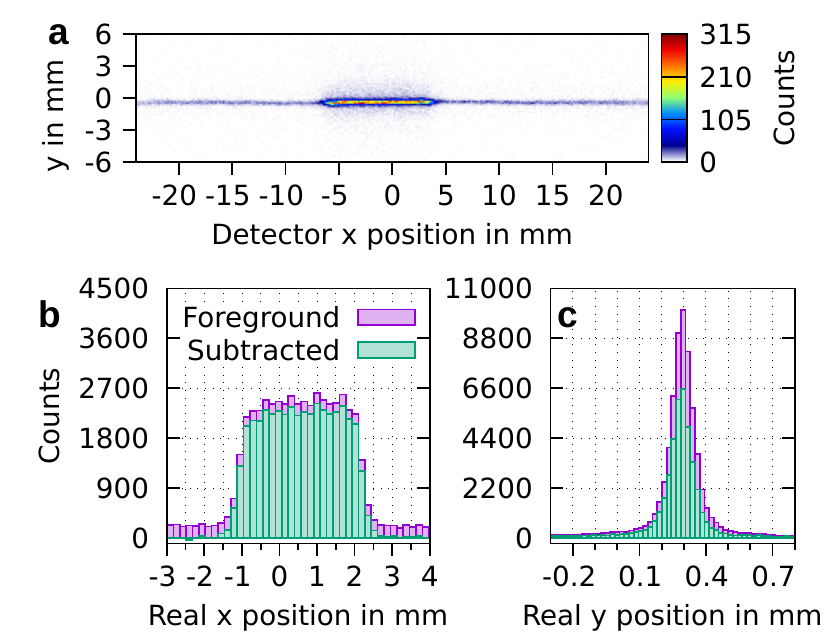}\\%
  \caption{\label{fig:smi}
    Cation spatial map image \textbf{(a)}, and its projections on the \ch{He}
    beam \textbf{(b)} and photon beam \textbf{(c)} axis. A \HND{} beam at
    \SI{30}{bar} backing pressure and \SI{12}{K} nozzle temperature is ionized
    by \SI{25}{eV} photons. Ionization of the background gas makes the photon beam visible and is
    subtracted from the projections to show the droplet related contribution.
    The conversion to real position is detailed in Supporting Information S5.
  }%
\end{figure}

For a cold \HND{} beam ionized at the synchrotron at \SI{25}{eV} photon energy,
the raw \SMI{} image on the detector and the centered projections in units of real length are
shown in Fig.~\ref{fig:smi}. The presented data are not mass selected, such that
the photon beam is clearly visible from ionized background gas.
The measurements were performed with a running chopper wheel. The rising and
falling edges of a \TTL{} monitor signal were acquired with the \TDC{}
to filter foreground and background events in the data analysis.
Fig.~\ref{fig:smi}b demonstrates the clear isolation of the droplet related
signal by subtraction.
The full width of the droplet beam profile at the baseline is \SI{4.3}{mm}, close
to the \SI{4}{mm} width of the scrapers that limit the horizontal extension of
the droplet beam.
The \FWHM{} of a Gaussian fit to the photon beam profile is
\SI{0.12}{mm} in good agreement with an estimate of \SI{0.11}{mm} derived from
ray-tracing results in the Supporting Information (S1).

\subsection{Time-of-flight mass spectrometry}

Currently two meshes separate the \VMI{} section from the \TOF{} section
which impedes to use the latter for velocity mapping.
In principle, the \TOF{} section can be used to collect electrons that
define a zero point in arrival time to perform mass selective cation
imaging in continuous beam experiments.
In case of electron imaging, the \TOF{} section provides complimentary
information on the mass of coincident cations which is a powerful tool to
probe if different ion formation channels involve different electronic
states.

\begin{figure}
  \includegraphics{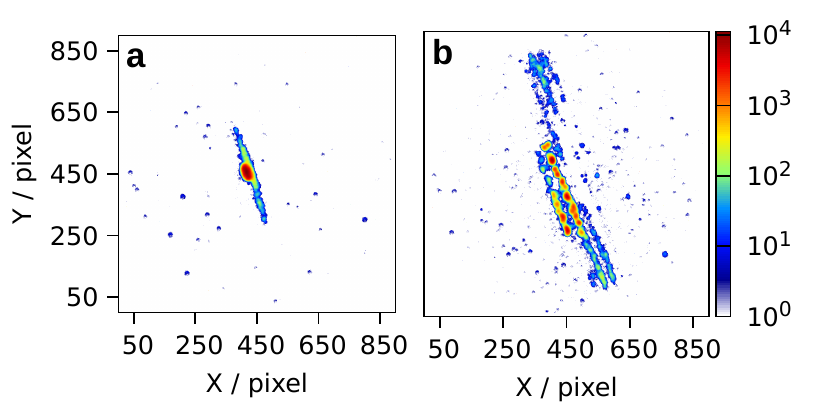}\\%
  \caption{\label{fig:defocus}
    Defocusing ions on the \TOF{} detector.
    Electrode potentials are
    E1: \SI{-1433.5}{V},
    R1/R2: \SI{-1750}{V},
    R5/R7: \SI{-3550}{V} and
    R6: \SI{-4250}{V}.
    The \MCP{} front and R6 potentials are the same.
    \textbf{(a)} Focused ions with $\mathrm{R3} = \mathrm{R2}$
    and $\mathrm{R4} = \mathrm{R5}$.
    \textbf{(b)} Defocused ions with
    $\mathrm{R3} = \mathrm{R4} = \SI{-3000}{V}$.
  }%
\end{figure}

It has been found that the specific potentials of the electrodes in the \TOF{}
section (see Fig.~\ref{fig:spectrometer}) have only a small effect on the time
resolution. Therefore, the potentials can be chosen to defocus the ions to
increase the detector lifetime. This is demonstrated by an integrated
intensity image of the phosphor screen in Fig.~\ref{fig:defocus}.

\begin{figure*}
  \includegraphics[width=\linewidth]{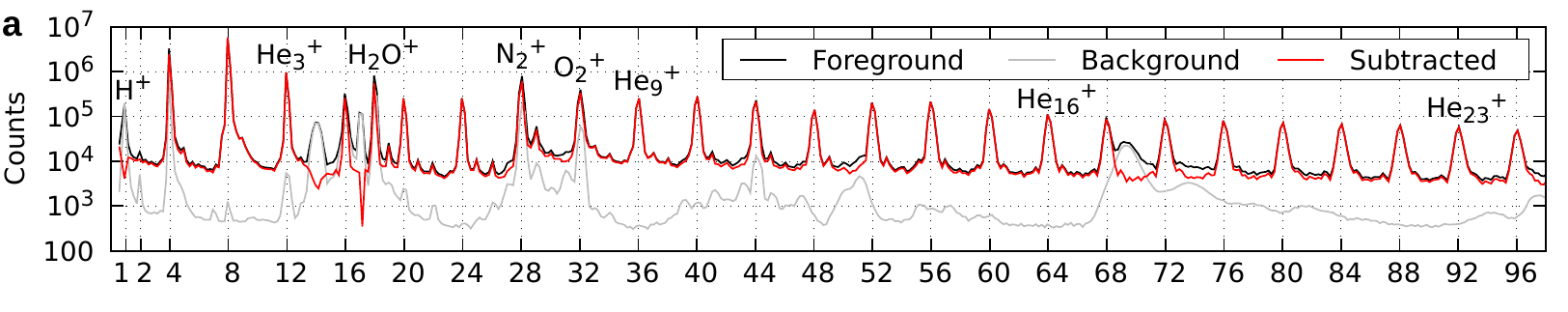}\\%
  \includegraphics[width=\linewidth]{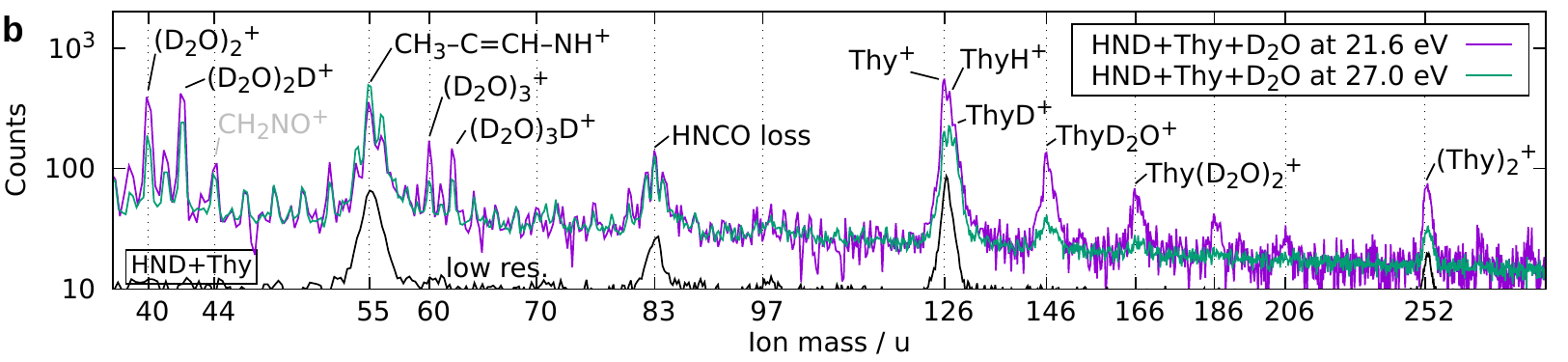}\\%
  \caption{\label{fig:massspec}
    Mass spectra of \textbf{(a)} pure \HND{}s ($h\nu=\SI{67.5}{eV}$)
    and \textbf{(b)} \HND{}s doped with thymine and \ch{D2O} where dopants are Penning
    ionized (\SI{21.6}{eV}) and charge-transfer ionized (\SI{27}{eV}).
    Acquisition times were about \SI{20}{\hour} (\SI{21.6}{eV})
    and \SI{6.5}{\hour} (\SI{27}{eV}) and the data of the latter are scaled by a
    factor of 0.3.
    The mass spectra in \textbf{(b)} are background subtracted.
    The solid black line shows a mass spectrum without \ch{D2O} and
    with reduced \TOF{} resolution at \SI{21.6}{eV} photon energy.
  }%
\end{figure*}

Two example mass spectra are presented in Fig.~\ref{fig:massspec}.
The first shows a progression of clusters from a pure \ch{He} beam.
The droplets were ionized
at \SI{67.5}{eV} above the first excited level of \ch{He+}.
Some \ch{He_{n}^{+}} peaks are mixed with water, nitrogen, oxygen and
carbon dioxide and their fragments from the background gas in the \VMI{} chamber
that also contains \ch{He+} from effusive \ch{He}.
Double electron coincidences from direct ionization and subsequent
\ICD{} at this photon energy are presented below.

The second mass spectrum shows the mass resolution achieved with the
\SI{100}{\percent} \VMI{} settings without using the Einzel lens electrodes.
For these data, \HND{}s were doped with the nucleobase thymine (oven
heated to \SI{100}{\celsius}, $\sim$ \SI{9.6e-6}{mbar} vapor pressure~\cite{Ferro1980:ta})
and deuterated water (\SI{3.3e-6}{mbar} partial pressure leaked directly into the vacuum chamber).
The nucleobase was ionized and fragmented by either Penning ionization from
\ch{He} excited at a photon energy $h\nu=\SI{21.6}{eV}$ or by charge transfer to \ch{He}
ionized at $h\nu=\SI{27}{eV}$. Single mass peaks are resolved up to around \SI{40}{u}, still clearly
visible around \SI{80}{u} and start to convolute around \SI{120}{u}.
This allows us to identify several thymine fragments with different hydrogen
content and mixed clusters with up to $4$ attached heavy water molecules. Also pure
water clusters up to $n=3$ are seen. The protonated water complex \ch{(D2O)_{3}D+}
is indicative for the presence of larger clusters.
Charge transfer leads to stronger fragmentation of thymine-\ch{D2O}
complexes and thymine clusters,
whereas in the case of Penning ionization,
larger clusters and thymine dimers are present in the spectrum.

\subsection{Energy scan of a Fano resonance}

Double excitation into the \ch{He} 2s2p state at $h\nu=\SI{60.15}{eV}$
gives rise to an atomic Fano resonance that is broadened and shifted to around
\SI{60.4}{eV} in \ch{He} nanodroplets~\cite{LaForge2016:pra}.
The autoionization of a doubly excited atom can compete with the ionization of
a neighbor atom. The second process is called Interatomic Coulombic Decay
(\ICD{}) and results into a lower electron kinetic energy (e\KE{}).
This has been theoretically shown for \ch{HeNe} dimers, where \ICD{}
is increasingly competitive with excitation into higher Rydberg
orbitals~\cite{Jabbari2020:cpl}. Unfortunately, investigating \ICD{} between
neighboring \ch{He} atoms in \HND{}s is complicated by
inelastic scattering of the photoelectron~\cite{Shcherbinin2019:jcp} giving
rise to the same e\KE{} if the same singly excited state is involved, e.g.\ 1s2p.
The inelastic scattering rate is expected to follow the Fano line shape of the
photoline. In that case the ratio between low and high e\KE{} would remain constant.
If, however, \ICD{} contributed to the electron yield, we would expect a pronounced increase of this ratio
at the double excitation resonance.

\begin{figure}
  \includegraphics{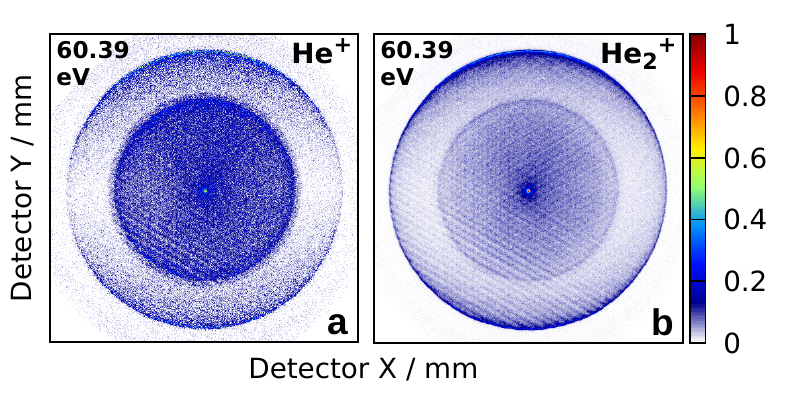}\\[-3pt]%
  \includegraphics[width=\linewidth,trim=8 0 10 5,clip]{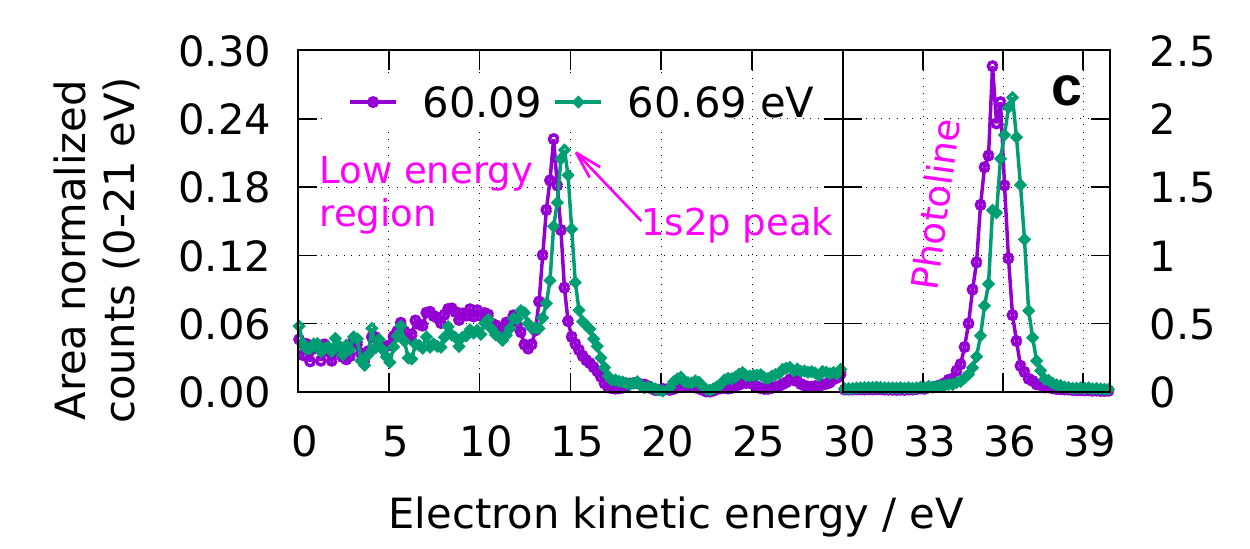}\\%
  \caption{\label{fig:fano}
    Electron images in coincidence with \textbf{(a)} \ch{He+} and \textbf{(b)}
    \ch{He2+} ions and \textbf{(c)} total e\KE{} distributions
    at different photon energies around the 2s2p \HND{} resonance
    (\SI{30}{bar} backing pressure, \SI{14}{K} nozzle).
    The \ch{He+} coincidence image and total e\KE{} spectra are
    background subtracted.
  }%
\end{figure}

We therefore performed detailed photon energy scans around the \SI{60.4}{eV} and
higher resonances that will be presented in a separate publication.
The \VMI{} potentials with a \SI{3.5}{kV} Einzel lens were ideally suited for
these measurements, providing excellent resolution and a wide electron
kinetic energy range
including photoelectrons (see Fig.~\ref{fig:nov:resolution}).
Fig.~\ref{fig:fano} shows an example of two electron images recorded in coincidence
with \ch{He+} and \ch{He2+} and the total e\KE{} distributions at different
photon energies. The photoline originates from direct ionization and autoionization while
the 1s2p peak is due to \ICD{} or inelastic scattering.
We can thus compare relative rates as a function of selected electron/ion
coincidences, electron kinetic energy and photon energy.

The \ch{He+} coincidence image in Fig.~\ref{fig:fano}a is background subtracted
to remove the contribution of background \ch{He} to the photoline (the \ch{He2+}
image is background free). In general, the background-subtraction factor is
corrected as described in the Supporting Information (S6).
To obtain energy distributions, foreground and
background images are inverted separately using \MEVELER{} to avoid pixel noise
and to fulfill the requirement of Poissonian statistics~\cite{Dick2014:pccp}.

\begin{figure}
  \includegraphics{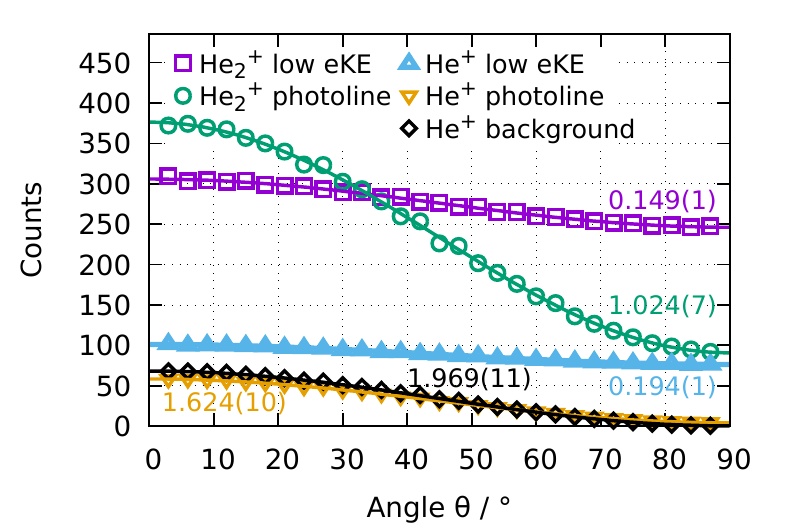}\\%
  \caption{\label{fig:fano:angle}
    Angle distributions at \SI{60.39}{eV} photon energy.
    Solid lines are
    $1 + \frac{\beta}{2} \left(3 \cos^2 \theta - 1\right)$ fits.
    Resulting $\beta$-values are shown in the same color.
    The curve for the \ch{He+} photoline is background
    subtracted.
  }%
\end{figure}

The electron images additionally contain angular information.
Distributions of the angle $\theta$ between the photoelectron velocity
vector and the polarization axis at \SI{60.39}{eV} photon energy are
presented in Fig.~\ref{fig:fano:angle} with fit results for the anisotropy
parameter $\beta$. The background-subtracted droplet contribution from the photoline in coincidence
with \ch{He+} slightly deviates from the nearly perfect $\cos^2\theta$
dependence as measured for background \ch{He} atoms. This indicates a deflection of
the emitted electrons presumably by elastic scattering of the
electrons at \ch{He} atoms. The \ch{He2+} photoline has a strong isotropic
contribution and the low-kinetic-energy electrons ($<\SI{22}{eV}$ in Fig.~\ref{fig:fano})
are predominantly isotropic due to massive elastic and inelastic
electron-He scattering. A detailed analysis of the electron angular
distributions of \ch{He} nanodroplets will be presented elsewhere.

\begin{figure}
  \includegraphics{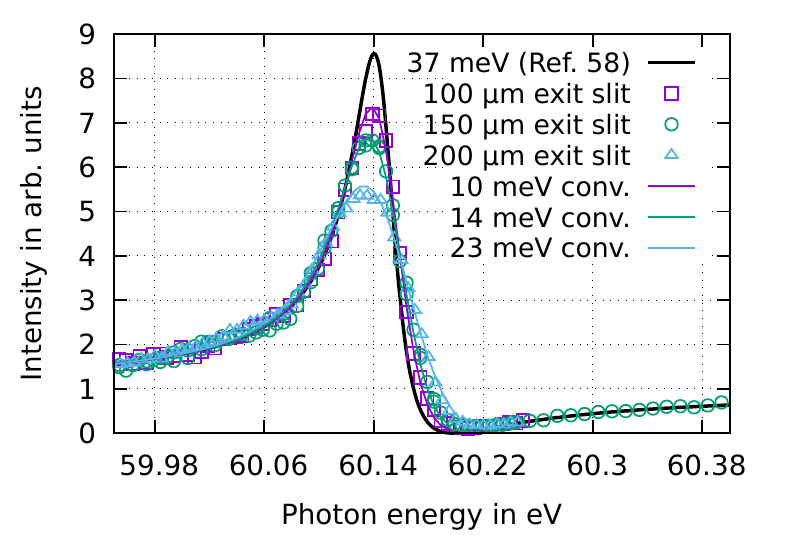}\\%
  \caption{\label{fig:fano:scan}
    Least-squares fits to an atomic Fano resonance for different monochromator
    exit slit widths. Gaussian functions with the specified widths are
    convoluted with the Fano profile (black curve) with parameters by
    Domke~\textit{et al.}~\cite{Domke1996:pra}
  }%
\end{figure}

The narrow lineshape of the atomic 2s2p Fano resonance allows us to determine
the energy resolution of the photon beam in the given energy range. To this end,
we convolute the Fano line shape~\cite{Kossmann1988:jpb} with a Gaussian function as described
by Domke \textit{et al.}~\cite{Domke1996:pra} The standard deviation of the Gaussian is determined by a least-squares
fit, see Fig.~\ref{fig:fano:scan}. The energy resolution depends on the width of the monochromator exit slit.
Measurements are usually taken with \SI{150}{\micro\meter} exit-slit width
corresponding to an absolute photon energy resolution of \SI{14}{meV} or a resolving power of 4300.
By further reducing the slit width to \SI{100}{\micro\meter} the resolution can
be further improved to \SI{10}{meV}, at the cost of reduced photon flux, though.
The measurement of the atomic Fano resonance position is also used for absolute energy
calibration of the beamline.

\subsection{Coincidences and multi-hits}
Owing to the multi-hit capability of the \HEX{} detector, it is possible to record two
electrons from the same ionization event. By identifying coincidences both
between electrons and between ions and electrons, we can select specific
ionization processes in the analysis. To demonstrate the capability of this
detection scheme, we revisit a previous experiment done at the GasPhase
beamline at Elettra~\cite{Shcherbinin2017:pra}. There, we studied Interatomic
Coulombic Decay (\ICD{}) in pure \ch{He} nanodroplets upon formation of an excited He
ion through the shake-up process.
\begin{multline}
\ch{He}-\ch{He} \xrightarrow{h\nu} \ch{He^{+*}} + \ch{He} + \ch{e_{sat}-} \\
\xrightarrow{\text{ICD}} \ch{He+} + \ch{He+} + \ch{e_{sat}-} + \ch{e_{ICD}-}
\end{multline}
The total number of charged particles from the process is then 4 (two electrons
and two cations). In our previous study of this process, detection of two ions
with different masses in coincidence with a single electron was possible, but
detection of multi-hits of electrons was impossible; only a single
electron was measured for each ionization event. With our new endstation,
detection of all products of the \ICD{} process is now possible. Furthermore,
with our optimized \VMI{} settings using the Einzel lens (see
section~\ref{sec:alternative:vmi}), we can image the electrons from direct
photoionization ($E_{\text{kin}} = h\nu - E_i^{\ch{He}}$) as well as the
satellite photoelectron (\ch{e_{sat}-}) from shake-up ionization and the
\ICD{} electron (\ch{e_{ICD}-}).

\begin{figure}
  \includegraphics[width=\linewidth]{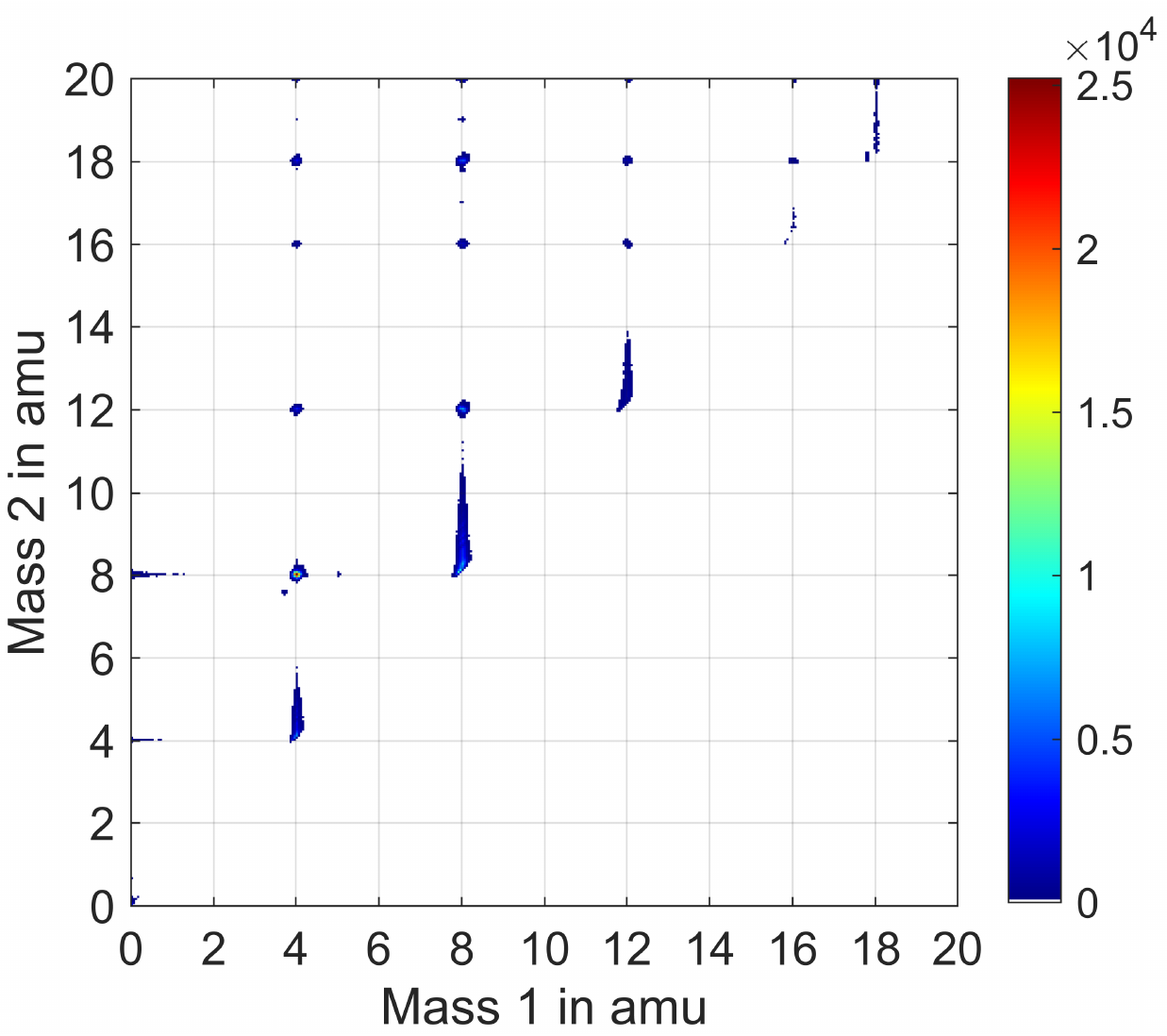}%
  \caption{\label{fig:shakeup:2d}
    Background-subtracted electron/ion/ion coincidence time-of-flight mass
    spectrum recorded at \SI{67.5}{eV} photon energy, \SI{30}{bar} backing
    pressure and \SI{14}{K} nozzle temperature.
  }%
\end{figure}

\begin{figure}
  \includegraphics{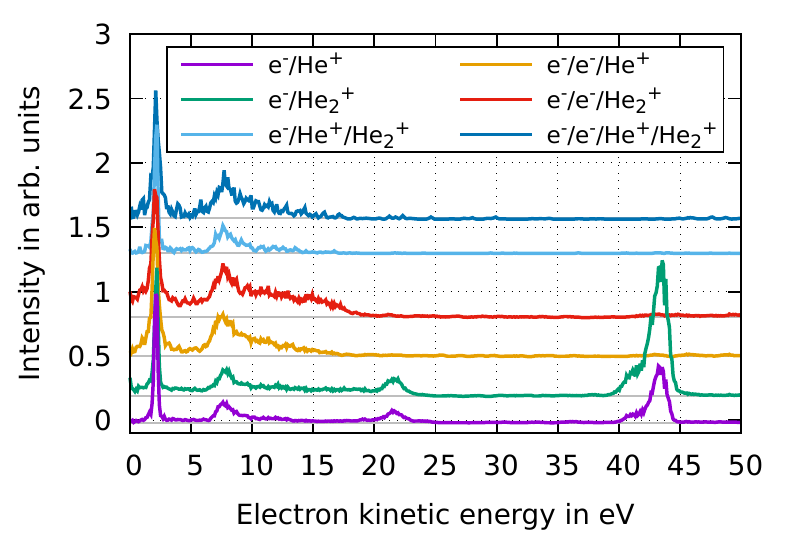}%
  \caption{\label{fig:shakeup:pes}
    Electron kinetic energy distribution for different coincidences recorded at
    \SI{67.5}{eV} photon energy (nozzle at \SI{14}{K} and \SI{30}{bar}).
    In the case of coincidences
    involving two electrons, the spectrum contains the sum of the two electrons.
  }%
\end{figure}

Figure~\ref{fig:shakeup:2d} displays coincidences between two ions recorded
at \SI{67.5}{eV}. Detection of two ions with identical mass-to-charge ratio is impossible
for the current use of the ion detector because they have nearly equal
flight times and thus one ion arrives within the deadtime of the detector caused by the other ion.
The most abundant ion-ion coincidence is the ion pair \ch{He+}/\ch{He2+}. Simultaneously detected two \ch{He_{$k$}+}
ions stem from either \ICD{} or impact ionization following direct
photoionization. The different ionization processes can be identified by
inspecting the electron kinetic energy spectra. Figure~\ref{fig:shakeup:pes}
displays electron spectra for various cuts on the coincidence data recorded at
\SI{67.5}{eV}. From the coincidences of a single electron with either the \ch{He+} or
\ch{He2+} ion, we can identify four peaks in the spectra. The peak at
$\sim\SI{43}{eV}$ is due to photoelectrons emitted by direct photoionization,
whereas the peak at $\sim\SI{22}{eV}$ is due to inelastic scattering of the photoelectron
with another \ch{He} atom in the droplet~\cite{Shcherbinin2019:jcp}.
In the case of direct photoionization in coincidence with \ch{He+}, the
photoline stems from the component of free atoms accompanying the droplet
beam.\cite{Shcherbinin2019:jcp}
The sharp line
at \SI{2.2}{eV} corresponds to satellite electrons, \ch{e_{sat}-}, and the broad feature centered
around \SI{8}{eV} are \ICD{} electrons, \ch{e_{ICD}-}.

Direct photoionization and inelastic electron-He scattering leading to
He excitation are processes creating only a single electron/cation pair.
Therefore, the two peaks from these processes are suppressed in the electron
spectra for coincidences of more than one electron and/or ion. The
\ch{e_{sat}-} and \ch{e_{ICD}-} electrons are present in the spectra
for all shown coincidence schemes, which shows that by selecting a certain
coincidence scheme, we can filter for a specific ionization channel\,---\,in this
case \ICD{}. The two electrons from the \ICD{} process arrive at the detector nearly
at the same time such that we cannot separate them into two images. Instead, the
spectra for coincidences of two electrons with one or two ions show the sum of
the two electrons from each event. Setting the condition that the two electrons
arrive at the detector simultaneously ($\Delta t < \SI{10}{ns}$) helps to
significantly suppress false coincidences. Differences in the shape of the \ICD{} feature are
seen depending on whether the coincidence cut includes the \ch{He+} ion or only
the \ch{He2+} ion. In triple \ch{e-}/\ch{e-}/\ch{He2+} coincidences, the
ICD feature extends to higher kinetic energy compared to the \ICD{} feature for
\ch{e-}/\ch{e-}/\ch{He+} coincidences. \ICD{} between two \ch{He} atoms at a larger interatomic
distance leads to an increased kinetic energy of the electron in connection
with a decreased kinetic energy release of the ions from Coulomb
explosion~\cite{sisourat2010ultralong}.
Thus, the high kinetic energy part in the \ICD{} electron peak for
\ch{e-}/\ch{e-}/\ch{He2+} coincidences is due to \ICD{} between two \ch{He} atoms at large
interatomic distance, which after ionization are more likely to pick up other \ch{He} atoms
before leaving the droplet due to their reduced kinetic energy. Consequently, the
detection of this high-energy part of the \ICD{} feature is suppressed for
\ch{e-}/\ch{e-}/\ch{He+} coincidences.

As already mentioned, electron impact ionization following direct photoionization can
also lead to two electron/cation-pairs. The characteristic feature for impact
ionization would be a broad feature reaching from zero kinetic energy to the
excess energy following impact ionization ($h\nu - 2 E_i^{\ch{He}} =
\SI{18.3}{eV}$) due to uniformly distributed energy sharing between the two
electrons~\cite{Shcherbinin2019:jcp}.
This is clearly seen in the \ch{e-}/\ch{e-}/\ch{He2+} coincidences
in agreement with the formation of \ch{He2+} after the creation of a slow ion
by electron impact.
In most other channels, a clear minimum of zero intensity is
present between the \ch{e_{sat}-} and \ch{e_{ICD}-} peaks, so we conclude
a minor importance of impact ionization.

From the detection of either a single electron or the two electrons from the \ICD{}
process, we can estimate the detection efficiency of electrons on the HEX75
detector using the following formulas
\begin{align}
  f_1  &= df_{\text{tot}} - f_2, \\
  f_2  &= d^2f_{\text{tot}},
\end{align}
where $f_{1,\,2}$ is the rate of detecting either one or two electrons per \ICD{}
event, $f_{\text{tot}}$ is the real \ICD{} event rate and $d$ is the detection
efficiency of each electron. The yields ($I_{1,\,2}$) of \ICD{} electrons from
observed one or two electron coincidences are then
\begin{align}
  I_1  &= f_1 t, \\
  I_2  &= 2f_2 t,
\end{align}
where $t$ is the measurement time, and the factor 2 comes from the fact that
electron spectra for coincidences of two electrons contain the sum of both
electrons. From the \ICD{}-related yield of coincidences including either a single
electron or both electrons, we then determine the detection efficiency to be
$\sim\SI{20}{\percent}$. From the $>$\SI{70}{\percent} open area ratio (\OAR{})
of the \MCP{}s, we would expect $\sim\SI{30}{\percent}$ loss of electrons at the
\MCP{}s~\cite{fehre2018absolute}. Further loss
of electrons happens at the mesh in front of the detector, and due
to dependencies of the detection efficiency on the kinetic energy of the
electrons~\cite{muller1986absolute}.

\subsection{Detection of anions}

Identification of ions by their time-of-flight in continuous experiments
relies on coincidence detection. The simplest case is photoionization, as
the electron flight time is negligible for mass spectrometry of the cations.
Anion detection is of interest for example to study electron attachment
processes.\cite{fabrikant2017recent} In \ch{He} nanodroplets, electrons with
relevant kinetic energies can be created by photoionization. After subsequent
resonant \cite{jaksch2009electron} or dissociative \cite{da2009electron}
electron attachment, the formed anion will be detected in coincidence with
the photoionized cation.

\begin{figure}
  \includegraphics[width=\linewidth]{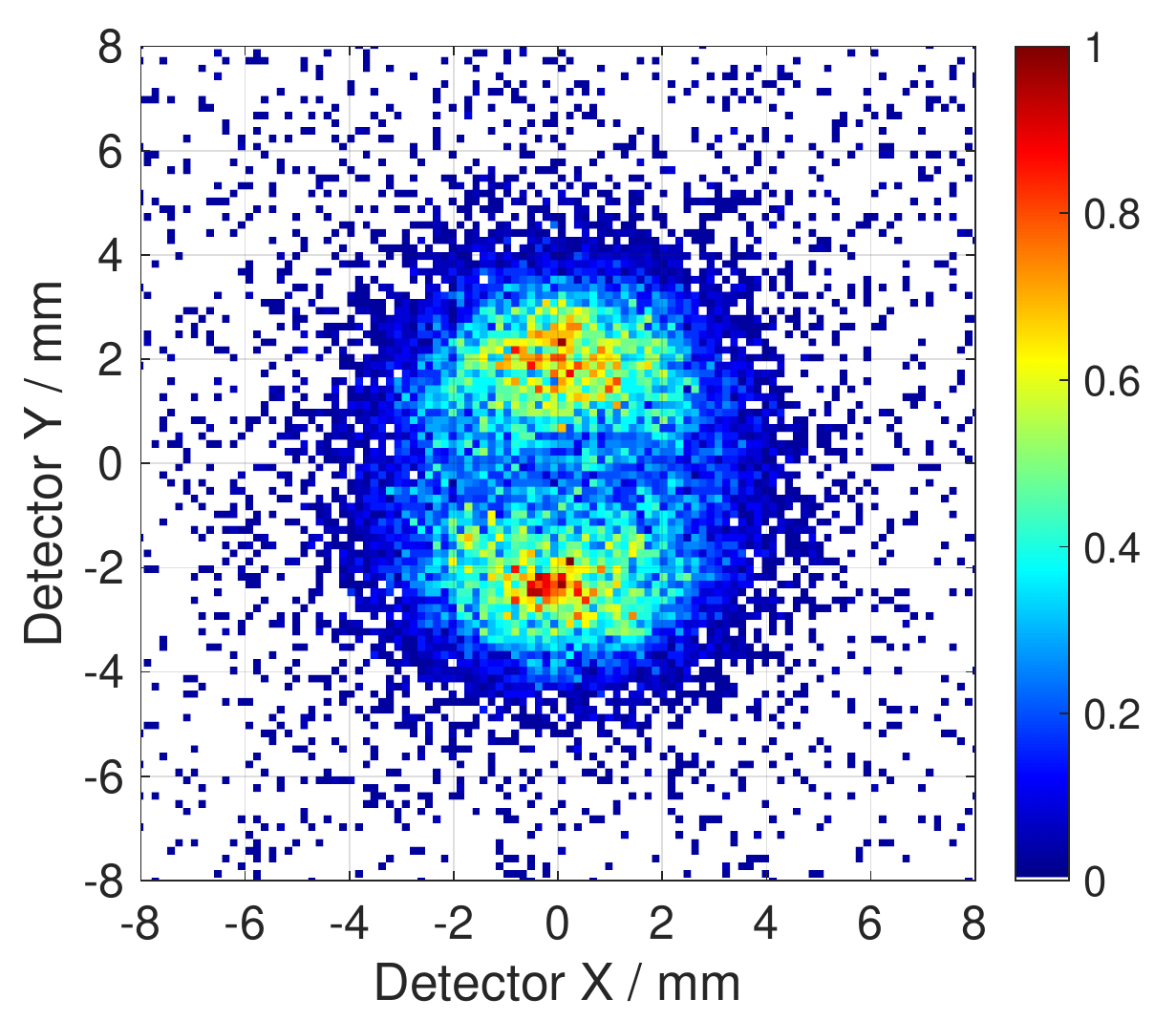}%
  \caption{\label{fig:anion:image}
    Ion images of \ch{O-} ions detected in coincidence with \ch{O+} following ion
    pair formation from \ch{O2} at $h\nu = \SI{17.47}{eV}$.
  }%
\end{figure}

As a proof of concept for anion detection, we probed resonant $\ch{O2} + h\nu
\rightarrow \ch{O+} + \ch{O-}$ pair formation at \SI{17.47}{eV}
photon energy~\cite{dehmer1975high,zhou2013ion}.
The cation flight time can be directly measured with photon energies above the
appearance energy of \ch{O+}. In case of pair formation, the event is triggered
by \ch{O-} on the \HEX{} detector and due to the longer flight time of the anion, the \ch{O+}
is observed at negative arrival time. The flight time of the coincident cation
plus the negative arrival time axis thus gives the anion \TOF{} spectrum and
allows to set a cut on the \ch{O-} image that is shown in Fig.~\ref{fig:anion:image}.

In such measurements of anion--cation coincidences, the primary
photoelectrons should be suppressed. For this purpose we have added a
simple construction to insert strong magnets near the ionization region.
While leaving the anion \TOF{} and image unperturbed, the electron count
rate was reduced by half. We expect to improve the suppression of electrons by adding a second
magnet tube on the opposite side.

\section{Conclusion and outlook}
In summary, we have described in detail a new endstation that complements the
experimental capabilities of the \AMOLine{} of the \ASTRID{} synchrotron radiation
facility. Using pure and doped \ch{He} nanodroplets that were resonantly excited or
ionized by \XUV{} radiation, we characterized the performance of the imaging
spectrometer with regard to high-resolution \VMI{} photoelectron spectra, ion
time-of-flight mass spectra, and spatial mapping of the ionization region.
The capability of detecting electrons and ions in coincidence up to
\ch{e-}/\ch{e-}/ion/ion quadruple coincidences was
demonstrated. Original multi-coincidence \ICD{} electron spectra in the regimes of
\ch{He}-droplet double excitation and shake-up ionization were discussed.
Furthermore, we presented measurements of oxygen anion/cation coincidences,
electron angular distributions of pure \ch{He} nanodroplets,
and fragment-ion mass spectra of thymine-\ch{D2O} complexes formed in \ch{He}
nanodroplets doped with thymine and water.

This endstation will be continuously upgraded in the following directions: A
continuous cluster source for producing heavier rare-gas clusters and water
clusters mixed with biomolecules is in preparation. The addition of an aerosol
injector is planned. Simultaneous imaging of electrons and ions in a
double imaging geometry~\cite{bodi2012new,garcia2013delicious} is envisioned.
The implementation of a \IR{}-\UV{} fluorescence spectrometer will further widen
the scope of future experimental studies.


%
%

%

\begin{acknowledgments}
The authors gratefully acknowledge financial support from Aarhus Universitets
Forskningsfond ({\small AUFF}) and from the Carlsberg Foundation. The authors are particularly
grateful for many insightful discussions with Robert Richter.
\end{acknowledgments}

\section*{Author declarations}

The authors have no conflicts to disclose.

\section*{Data Availability Statement}

The data that support the findings of this study are available from the corresponding author upon reasonable request.

%

\end{document}


\setstretch{1}


\title{%
  \texorpdfstring{%
    \LARGE A new endstation for extreme-ultraviolet spectroscopy of free
    clusters and nanodroplets \\[12pt]
    \Large Supporting Information}{%
    A new endstation for extreme-ultraviolet spectroscopy of free clusters
    and nanodroplets: Supporting Information}}

\setstretch{1.3}



\author{\vspace{12pt}Bj\"{o}rn Bastian}
\author{Jakob D. Asmussen}
\author{Ltaief Ben Ltaief}
\affiliation{Department of Physics and Astronomy,
  Aarhus University, Ny Munkegade 120, 8000 Aarhus C, Denmark}
\author{Achim Czasch}
\affiliation{Institut für Kernphysik, Goethe Universität,
  Max-von-Laue-Strasse 1, 60438 Frankfurt, Germany}
\author{Nykola Jones}
\author{S{\o}ren V. Hoffmann}
\author{Henrik B. Pedersen}
\author{Marcel Mudrich\vspace{6pt}}
\affiliation{Department of Physics and Astronomy,
  Aarhus University, Ny Munkegade 120, 8000 Aarhus C, Denmark}
\email{mudrich@phys.au.dk}
\noaffiliation

\date{\today}

\begin{abstract}
\end{abstract}

\pacs{}

\maketitle 

\section{Refocusing optics for the endstation}
\enlargethispage{\baselineskip}

\begin{figure*}[ht]
  \includegraphics[width=.65\linewidth]{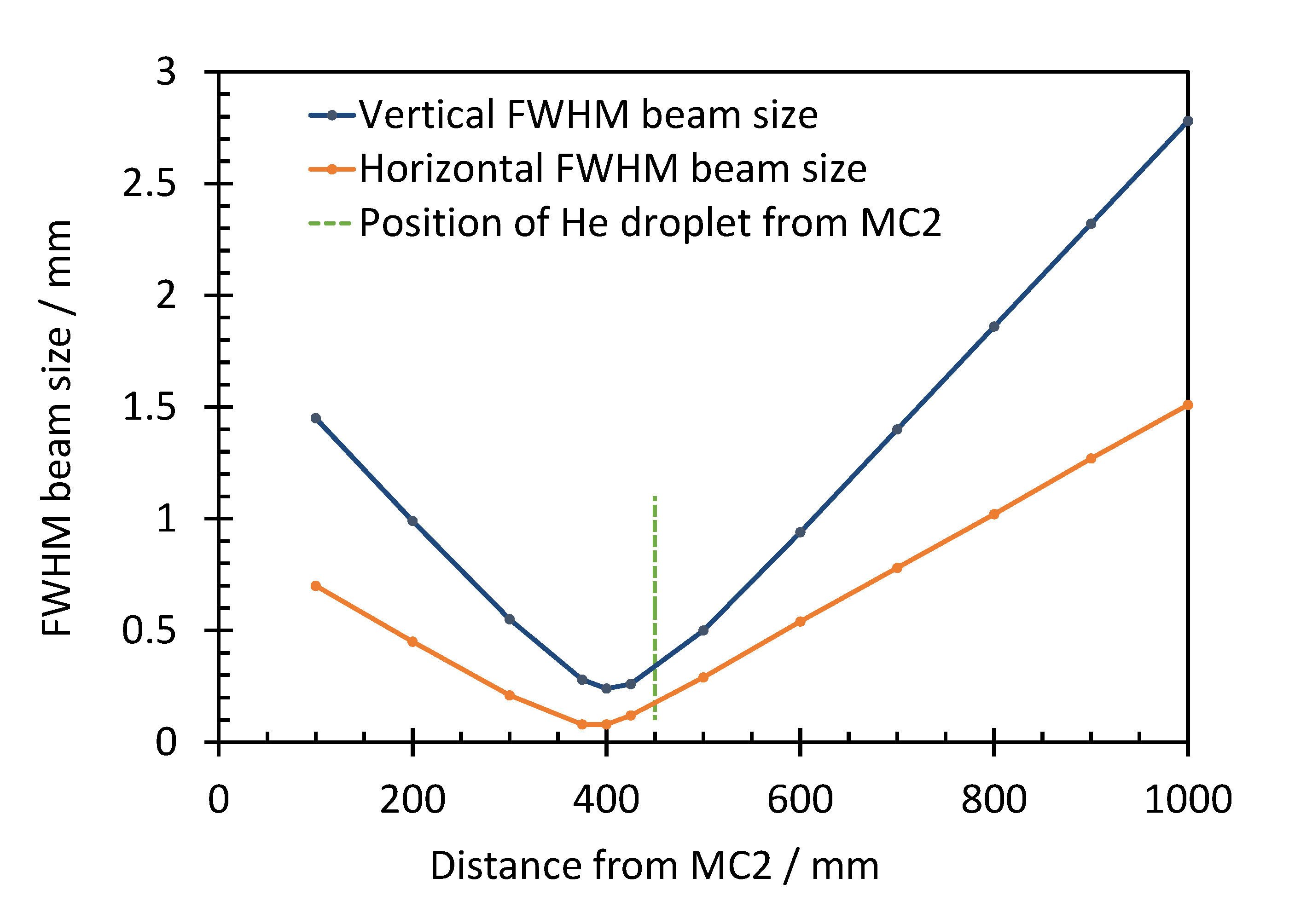}
  \caption{\label{fig:mirror}%
    Beam size of the refocused photon beam as a function 
    of the distance from the mirror.
  }
\end{figure*}

The new endstation for extreme-ultraviolet spectroscopy of free clusters and
nanodroplets is placed at the end of the existing \ASTRID{} synchrotron
radiation beam line, the \AMOLine{}.
To refocus the photon beam from the beam line into the sample region of the
endstation, a toroidal mirror, \MCTWO{}, is placed between the \AMOLine{} and
the endstation.
The \MCTWO{} mirror images an intermediate focus (\IF{}) in the \AMOLine{} such
that the distance from the \IF{} to \MCTWO{} is \SI{1080}{mm} and the distance
from \MCTWO{} to the geometrical focus is \SI{500}{mm} under a reflection angle
of 88 degrees from the mirror surface normal.
The mirror parameters are shown in Tables~\ref{tab:geometry} and
\ref{tab:substrate}.
Since the mirror is operated under a very grazing angle with the incoming beam
2 degrees from the mirror surface, and as the focus distance from the mirror is
relatively short compared to the length of the mirror, the distance from the
mirror where the beam size is smallest is shorter than the geometrical focus
distance.
To investigate the beam size as a function of the distance from the refocusing
mirror, ray-tracing was done with Shadow\cite{Lai1986:nima} using a simulated
undulator source and including all the optical elements in the AMOLine{}.
The results are shown in Fig.~\ref{fig:mirror}, and the minimum beam size is
close to \SI{400}{mm} from the mirror.
When using the first harmonic of the undulator, the photon source is a single
beam having a close to Gaussian shape, whereas the third harmonic of the
undulator would have an angular distribution with a central cone and a ring of
light at larger angles.
To ensure that the use of the mirror at a distance of \SI{400}{mm} from
\MCTWO{}, i.e.\ away from the geometrical focus, does not induce a multi-spot
image of a third harmonic undulator source, both the first (\SI{10}{eV})
and the third harmonics of the undulator were ray-traced.
They showed very similar sized and single-spot images.
Therefore, based on these simulations, a distance of less than \SI{500}{mm} from the
\MCTWO{} mirror to the droplet beam can be used for the placement of the
endstation resulting in smaller beam size. Due to space constraints in the
experimental setup a final distance of \SI{450}{mm} was chosen.

\begin{table*}[ht]
  \caption{\label{tab:geometry}%
    Geometry for the use of the mirror.
  }
  \begin{tabular}{l@{~~}r}
    \toprule
    Object to mirror distance ($r$) & \SI{1080}{mm} \\
    Mirror to Image distance ($r'$) & \SI{500}{mm} \\
    Reflection angle to mirror normal ($\alpha$) &
    \SI{88}{\degree}~\quad\hbox{\,} \\
    \hline
  \end{tabular}
  \caption{\label{tab:substrate}%
    Substrate shape parameters.
  }
  \begin{tabular}{l@{~~}l}
    \toprule
    Shape type & Toroidal \\
    Tangential radius ($R$)  & \SI{19600}{mm} (along long side of the mirror) \\
    Sagittal radius ($\rho$) & \SI{23.85}{mm} (along short side of the mirror) \\
    Optical active area & $\SI{55}{mm} \times \SI{10}{mm}$ \\
    \hline
  \end{tabular}
\end{table*}

For comparison with spatial map imaging results we estimate the horizontal beam
size at the position of the \ch{He} droplet beam for \SI{25}{eV} photon energy.
The ray-tracing at \SI{10}{eV} photon energy yields ca.\ \SI{0.18}{mm} \FWHM{}
in Fig.~\ref{fig:mirror}.
The diffraction limited beam size
$\sigma_r \propto \sqrt{\lambda/L}$
with wavelength $\lambda$ and undulator length $L$ decreases with energy, and
we thus expect a \FWHM{} of \SI{0.11}{mm} at \SI{25}{eV} photon energy.

\section{Numerical optimization for velocity map imaging\label{app:alternative:vmi}}

To determine extractor electrode potentials for precise
\VMI{} without using the Einzel lens, the repeller R1 and R2 potentials in
Fig.~3 
(main article) are set to \SI{-3500}{V}. The maximum potential
from the power supplies is \SI{-6}{kV}, so the chosen repeller potentials
allow for cation acceleration in the \TOF{} section up to \SI{2500}{eV} kinetic
energy which is more than sufficient for single ion detection with a \MCP{}.
The potentials of the three extractor electrodes E1--E3 are then adjusted to
obtain optimal resolution. The Einzel lens potentials E4--E6 are kept at
ground potential and the \MCP{} front potential is \SI{240}{V} for electron
imaging.
Using the \SIMION{}-based \VMI{} optimization routine published by Stei~\emph{et al.}\cite{Stei2013:jcp}
for an ion at rest in the center of the spectrometer yields the potentials
$\mathrm{E1} = -\SI{2867}{V}$, $\mathrm{E2} = -\SI{981}{V}$ and $\mathrm{E3} =
-\SI{391}{V}$. The routine is based on computing the first and second order
derivatives of impact position and time on the detector with respect to the
initial position and velocity vectors. To optimize for two-dimensional \VMI{},
the electrode potentials are then varied to optimize the target function
\begin{align}
  \label{eq:target}
  \left|\frac{\partial R}{\partial r}\right|
  + \left|\frac{\partial R}{\partial z}\right|
  + \left(\left|\frac{\partial R}{\partial v_r}\right|\right)^{-1}
  + \left|\frac{\partial R}{\partial v_z}\right|
  + \sum_{i,j=1}^{4} \left|\frac{\partial^2 R}{\partial x_i \partial x_j}\right|
\end{align}
with $\{x_i\} = \{r, z, v_r, v_z\}$ where
$R$ denotes the detector radius at the impact
position and the $x_i$ the initial position $r$, $z$ and velocities $v_r$,
$v_z$ in cylinder coordinates. The $z$-axis is parallel to the symmetry axis
of the spectrometer.
The target function could be refined by explicitly taking into account the
initial spatial broadening and velocity spread of the ions, which however has
its limits by the enforced rotational symmetry about the detector axis and the
dependence of the velocity distribution on the probed process.
Instead, we follow the generic procedure by Stei~\emph{et al.}: The choice of
native units with all quantities on the order of magnitude $1$ allows to set
all prefactors to $1$.
Therefore, Eq.~\eqref{eq:target} mixes different units and dimensions but is
able to roughly homogeneously suppress the different factors that otherwise
contribute to broadening of the desired velocity mapped distribution.

The simplest approach to increase the kinetic energy acceptance is the scaling
of a set of \VMI{} potentials to reduce the impact radii on the detector.
The maximum available potential with the current power supplies is \SI{6}{kV}.
Increasing the repeller potential is thus limited to about \SI{4}{kV} by the
requirement of about
\SI{2}{kV} acceleration for ion detection in case of coincidence measurements.
Instead, up to \SI{90}{eV} kinetic energy acceptance is achieved by using the
Einzel lens electrodes with the central electrode on ground (close to the \MCP{}
front potential) and the outer electrodes on high positive potentials to
temporarily accelerate the electrons and thus compress the spatial distribution
on the detector.
To determine \VMI{} potentials with the Einzel lens, the previously
optimized potentials were modified by adding a fixed potential to the Einzel
lens electrodes E4 and E6 (see Fig.~3 
in the main article)
and then reoptimized for the extractor potentials E1, E2 and E3 to restore good
\VMI{} conditions, that is, a small dependence from the initial position of the
imaged particle.
In contrast to the originally published optimization routine,
the first and second order derivatives were typically evaluated at non-zero
initial kinetic energies as specified for the different optimization results in
Table~\ref{tab:vmi:alt}.

\begin{table}
  \caption{\label{tab:vmi:alt}%
    Alternative \VMI{} potentials in \si{V}. Settings with lens use the Einzel lens
    electrodes with $\mathrm{E4} = \mathrm{E6}$ and $\mathrm{E5} = \SI{0}{V}$.
    Otherwise $\mathrm{E5} = \mathrm{E6} = \SI{0}{V}$.
    (1), (2) and (3) denote numerical optimization for \SI{0.5}{eV},
    \SI{30}{eV} and \SI{70}{eV} kinetic energy ions.
    The linear calibration coefficient $p_0$ and the ratio between its simulated
    and experimental value are given in the last two columns.
  }
  \begin{tabular}{l@{~\,}c@{~\,}c@{~\,}c@{~\,}c@{~\,}c@{~\,}l@{~\,}l}
    \toprule
    Settings & R1    & E1    & E2    & E3    & E4     & ~~Exp. $p_0$ & Sim./Exp. \\
    \hline
    \SI{3.5}{kV} lens {\small (3)} & -3500 & -2628 & -1359 & 71 & ~3500 & \num{0.0331(3)}   & \quad 1.09 \\
    \SI{6.0}{kV} lens {\small (3)} & -3500 & -2634 & -1152 & -388 & ~6000 & \num{0.0505(3)}   & \quad 1.37 \\
    Non-opt.\ lens & -3500 & -2867 &  -981 &  -391 & ~6000 & \num{0.0602(3)}   & \quad 1.06 \\
    No lens \SI{50}{\percent} & -1750 & -1433.5 &  -490.5 &  -195.5 &     0 & \num{0.01222(7)}  & \quad 0.96 \\
    No lens & -3500 & -2867 &  -981 &  -391 &     0 & \num{0.02348(14)} & \quad 1.004 \\
    ~with E4 {\small (1)} & -3500 & -3240 & -2935 & -2397 & -1308 \\
    ~with E4 {\small (2)} & -3500 & -3236 & -2929 & -2419 & -1380 & \num{0.0237(2)}   & \quad 1.03 \\
    \hline
  \end{tabular}
\end{table}

Simulations have shown that evaluating the derivatives for the numerical
optimization at different initial kinetic energies
yields \VMI{} potentials that have the optimal spatial focusing at
different detector radii. This has not found to be significant in the empiric
characterizations, which has been explained by the spatial resolution of the
detector system that determines the energy resolution at low detector radii.
Optimizations at non-zero kinetic energy are still useful to obtain potentials
that are theoretically less sensitive to the initial position and with
calibration functions that deviate less from linearity.

All optimization results were obtained for an initial design that included a
mesh at the exit side of electrode E6. Without the mesh, the magnification
is reduced and the initial kinetic energy of \SI{70}{eV} (see caption to
Table~\ref{tab:vmi:alt}) corresponds to a new value of \SI{59}{eV} with the same
impact radius on the detector. Potentials without the Einzel lens do not
depend on the presence of the mesh. Reoptimizations without the mesh for
\VMI{} settings with the Einzel lens give similar results and are expected to
also be comparable to the presented results with respect to energy resolution.

For the standard potentials without compression, the first electrode E4
of the Einzel lens section can be used to further improve the \VMI{}
focusing conditions. Two such settings are listed in Table~\ref{tab:vmi:alt}.
The weaker extraction potential however reduces the time of flight resolution.
It has not been tested if optimization at nonzero kinetic energies could
further improve the resolution of the \SI{100}{\percent} potentials without
E4 and keeping the steeper extraction potential.
The potentials with E4 from optimization at \SI{0.5}{eV} kinetic energy
were systematically varied to check, if the optimization results match optimal
empirical results in terms of energy resolution.
Only the E1 potential turned out to be important
and matched the optimum up to \SI{20}{V} precision.
Variations of the E1 potential could not improve the overall energy resolution
in calibration measurements in a significant way.

The measured and simulated linear calibration factors $p_0$ are given in the
last columns of Table~\ref{tab:vmi:alt}. Discrepancies are on the order of
\SI{5}{\percent} or smaller, but increase in case of the optimized settings
with high Einzel lens potentials up to \num{9} and \SI{37}{\percent}.

\section{\label{app:calibration}Energy calibration for velocity map imaging}

\VMI{} maps an electron moving perpendicular to the spectrometer axis (slice
distribution) with a specific kinetic energy $E_\mathrm{kin}$ onto a specific
radius $R$ on the detector.
A three-dimensional momentum vector distribution with cylindrical symmetry
results in an Abel transformed image on the detector such that the slice
distribution can be reconstructed by Abel inversion.
Assuming an analytical dependence of the kinetic energy on the radius and
symmetry about the spectrometer axis, only even powers in a Taylor
expansion are allowed, which in $3^{rd}$ order gives
\begin{equation}
  \label{eq:calibration}
  E_\mathrm{ph} = E_0 + p_0 \left(R^2\right) + p_1 \left(R^2\right)^2 + p_2 \left(R^2\right)^3
\end{equation}
for the photon energy, where $E_0$ is the ionization potential of helium.
For each photon energy, the radius of impact on the detector $R$ is obtained
from a Gaussian fit to the radial distribution of the Abel inverted image.
The inversion is performed with the Maximum Entropy Velocity Legendre
Reconstruction (\MEVELER{}) method \cite{Dick2014:pccp}.
Fitting Eq.~\eqref{eq:calibration} to the results in Fig.~5a 
yields the coefficients in Table~II 
with residuals $<\SI{1}{\permil}$.
Simulation results were obtained for centered ions with velocity vectors
perpendicular to the spectrometer axis without an inversion step.

The calibration data are also useful to determine the resolution $\Delta E / E$
as a function of kinetic energy.
All calibration and energy resolution related data
were measured in a row. This allows reliable
comparisons of different \VMI{} potentials with otherwise identical
conditions. The energy resolution could not always be reproduced between
different beamtimes.

\section{\label{app:resolution}Energy resolution with velocity map imaging}

To assess the \VMI{} energy resolution from simulations, detector images have
been computed for groups of \num{15000} particles with the same kinetic energy.
The initial velocity direction and positions were uniformly sampled from a
$4\pi$ solid angle and a cylinder with \SI{4}{mm} length, corresponding to the
width of the \ch{He} droplet beam (along the photon beam), and a radius of
\SI{0.2}{mm} of the refocused photon beam, corresponding to \num{4} standard
deviations of its radial profile. The spatial distribution of the impact
positions was binned in $\SI{0.1}{mm} \times \SI{0.1}{mm}$ squares and Abel
inverted.  The energy resolution was obtained from the standard deviation and
position of a Gaussian fit to the resulting radial distribution according to
Gaussian error propagation as $\Delta E / E = 2~\Delta r / r$ with the radius
$r$ on the detector.

\section{Spatial map imaging\label{app:smi}}

The spectrometer can be operated at different potentials to image positions in
the interaction region with high precision\cite{Stei2013:jcp} instead of
performing velocity map imaging.
Table~\ref{tab:smi} summarizes electrode potentials before and after
optimization for spatial map imaging (\SMI{}) along with the relation between
the radial impact position on the
detector $R$ and the initial position $r$ and radial velocity $v_r$ of the ion.
While the magnification factor $\partial R/\partial r$ stays similar, the effect
of the initial velocity, in first order $\partial R / \partial v_r$, is
substantially reduced.
This effect could be confirmed by the observation of sharper \SMI{} images.

\begin{table}
  \begin{minipage}[t]{.48\linewidth}
    \caption{\label{tab:smi}%
      \setstretch{1.03}
      Empirical and simulated \SMI{} potentials in \si{V} and first
      derivatives of the impact position $R$ with respect to initial $r$, $v_r$.
      Space and velocity units are \si{mm} and \si{\milli\meter\per\micro\second}.
    }
    \small
    \begin{tabular}{lcccccc}
      \toprule
      Settings & R1 & E1 & E2 & E3 & $\partial R/\partial r$  & $\partial R/\partial v_r$ \\
      \hline
      Empirical & 3500 & 3495 & 3230 & 0   & -3.395 &  0.256 \\
      Optimized & 3500 & 3469 & 3456 & 213 & -3.154 & -0.001 \\
      \hline
    \end{tabular}
  \end{minipage}
  \hfill
  \begin{minipage}[t]{.48\linewidth}
    \caption{\label{tab:smi:alt}%
      Alternative \SMI{} potentials \hbox{in \si{V}.}
      Magnifying potentials map radial positions $r$
      with $\partial R/\partial r = \num{-13.34}$
      and \TOF{}-\SMI{} maps axial positions $z$ with
      $\partial T/\partial z = \SI{58.3}{\nano\second\per\milli\meter}$.
    }
    \begin{tabular}{lcccccc}
      \toprule
      Settings & R1 & E1 & E2 & E3 & E4 & E5 \\
      \hline
      Magnified     & 3500 & -2939 & 2726 & -1106 & 0 & 262 \\
      \TOF{}-\SMI{} & 3500 & 1073 & -3498 & 334   & 680 & 0 \\
      \hline
    \end{tabular}
  \end{minipage}
\end{table}

With different initial potentials or target functions in the optimization,
alternative settings were obtained to achieve larger magnification or to perform
one-dimensional \SMI{} using the ion time of flight.
The potentials are given in Table~\ref{tab:smi:alt}.
The simulated magnification of \num{-13.34} is about four times larger than
before but at the cost of a significant derivative
$\partial R/\partial v_r = \SI{3.28}{\micro\second}$.
The scaling factor for the \TOF{}-\SMI{} is
\SI{58.3}{\nano\second\per\milli\meter} and the first order influence of
the ion velocity on the arrival time is
$\partial T/\partial v_z = \SI{-1.8}{\nano\second\micro\second\per\milli\meter}$.

Simulated and measured magnification factors for electron and cation \SMI{} as
well as widths of the spatial distributions are given in
Table~\ref{table:smi:calibration}. They are discussed with regard to the
achieved \SMI{} resolution in the main article.

\begin{table*}[ht]
  \caption{\label{table:smi:calibration}%
    \SMI{} magnification factors from \ch{K}-doped \HND{} ionized at \SI{767}{nm}.
    Empirical and simulated values as well as the relative deviation and widths of
    the spatial distributions orthogonal to the ionization laser are given for
    different \SMI{} potentials and for cation and electron imaging.
  }
  \begin{tabular}{lc@{~~}c@{~~}c@{~~}c@{~~}c}
    \toprule
    SMI setting & Species & Empiric factor & Simulated factor & Deviation
    & Width/mm \\
    \hline
    SMI 1D & \ch{K+} & \SI{64.0}{ns/mm} & \SI{58.3}{ns/mm} & \SI{10}{\percent}\,~  & 0.050 \\ 
    SMI 2D (magn.) & \ch{K+} & -13.0~~~~~ & -13.34 & ~~\,\SI{2.6}{\percent} & 0.050 \\ 
    SMI 2D (\SI{100}{\percent}) & \ch{K+} & -3.200 & -3.154 & \SI{-1.5}{\percent} & 0.050 \\ 
    SMI 2D (\SI{100}{\percent}) & \ch{e-} & -3.203 & -3.154 & \SI{-1.6}{\percent} & 0.08~ \\ 
    SMI 2D (\SI{50}{\percent}) & \ch{e-} & -3.202 & -3.154 & \SI{-1.5}{\percent} & 0.11~ \\ 
    \hline
  \end{tabular}
\end{table*}

The real positions in Fig.~8 from \SMI{} of \ch{He+} in the main article are
determined with the calibration factor \num{-3.2} and with respect to the zero
point $x_0 = 0.43$, $y_0 = 0.5$. The zero point of the spectrometer was
determined as the center of the photoelectron rings in an electron \VMI{}
calibration measurement of an atomic \ch{He} beam.
The systematic error of the zero point determination due to the center of mass
velocity of the \ch{He} beam should be small because of the low electron mass
and consequently short flight time to the detector.
Using the estimate of \SI{560}{\meter\per\second} for the maximum speed
of an ideal supersonic expansion of \ch{He} at \SI{30}{K},
a \SIMION{} trajectory yields a deviation of \SI{5}{\micro\meter}.


\section{Correction of the background subtraction factor}

For total electron distributions, the background is subtracted from the
foreground with a weight of 1, if the acquisition durations are identical.
Conditions to the data that restrict the hit statistics, for instance
to analyze only single hits, are typically more restrictive to the foreground
than to the lower count rate background data and require the correction
\begin{align}
  I_\mathrm{subtracted}
  = I_\mathrm{fg} - \frac{ T_\mathrm{fg} }{ T_\mathrm{bg} }
  \frac{ N_\mathrm{bg, all} }{ N_\mathrm{fg, all} }
  \frac{ N_\mathrm{fg, restr} }{ N_\mathrm{bg, restr} }
  I_\mathrm{bg}
\end{align}
with time intervals $T_\mathrm{x}$ of the chopper wheel period that are
collected into foreground ($\mathrm{x}=\mathrm{fg}$) and background
($\mathrm{x}=\mathrm{bg}$) data, the total electron
counts $N_\mathrm{x, all}$ and the restricted electron counts
$N_\mathrm{x, restr}$.

\setstretch{1}